\PassOptionsToPackage{dvipsnames,usenames}{xcolor}

\documentclass[conference]{IEEEtran}

\AtBeginDocument{%
  \providecommand\BibTeX{{%
    \normalfont B\kern-0.5em{\scshape i\kern-0.25em b}\kern-0.8em\TeX}}}

\usepackage{amsmath,amssymb,amsfonts}
\usepackage{multirow}
\usepackage{hhline}
\usepackage{balance}
\usepackage[font=small,labelfont=bf]{caption}
\usepackage{algorithm}
\usepackage[noend]{algpseudocode}
\usepackage[hidelinks]{hyperref}
\usepackage{cleveref}
\usepackage[htt]{hyphenat}
\usepackage{float}
\usepackage{bm}
\usepackage{xspace}
\usepackage{subcaption}
\usepackage{enumitem}
\usepackage[commandnameprefix=always]{changes}

\newcommand*{\costream}{\textsc{Costream}\xspace}
\newcommand*{\flatvector}{\textsc{Flat Vector}}

\newcommand*\circles[1]{\raisebox{.5pt}{\textcircled{\raisebox{-.8pt} {#1}}}}

\newtheorem{definition}{Definition}
\usepackage[acronym,toc,shortcuts]{glossaries}
\newacronym{dsps}{DSPS}{Distributed Stream Processing Systems}
\newacronym{dag}{DAG}{Directed Acyclic Graph}
\newacronym{mlps}{mlps}{Multi-Layer Perceptrons}
\newacronym{qos}{QoS}{Quality of Service}
\newacronym{iot}{IoT}{Internet of Things}
\newacronym{ml}{ML}{Machine Learning}
\newacronym{dcs}{DCs}{Data Characteristics}
\newacronym{gnn}{GNN}{Graph Neural Networks}
\usepackage[numbers,sort&compress]{natbib} 

\begin{document}

\title{\costream: Learned Cost Models for Operator Placement in Edge-Cloud Environments}

\author{
    \IEEEauthorblockN{Roman Heinrich}
    \IEEEauthorblockA{DHBW Mannheim}
    \and
    \IEEEauthorblockN{Carsten Binnig}
    \IEEEauthorblockA{TU Darmstadt \& DFKI}
    \and
    \IEEEauthorblockN{Harald Kornmayer}
    \IEEEauthorblockA{DHBW Mannheim}
    \and
    \IEEEauthorblockN{Manisha Luthra}
    \IEEEauthorblockA{TU Darmstadt \& DFKI}
}

\maketitle

\begin{abstract}
In this work, we present \costream, a novel learned cost model for Distributed Stream Processing Systems that provides accurate predictions of the execution costs of a streaming query in an edge-cloud environment.
The cost model can be used to find an initial placement of operators across heterogeneous hardware, which is particularly important in these environments.
In our evaluation, we demonstrate that \costream{} can produce highly accurate cost estimates for the initial operator placement and even generalize to \textit{unseen} placements, queries, and hardware.
When using \costream{} to optimize the placements of streaming operators, a median speed-up of around $21\times$ can be achieved compared to baselines. 
\end{abstract}

\vspace{-0.5ex}
\section{Introduction} \label{sec:introduction}
\vspace{-1ex}
\textbf{Operator placement in distributed stream processing.} 
\ac{dsps} play a crucial role in a wide spectrum of high-performance applications, enabling efficient and scalable processing of unbounded data streams.
Therefore, these systems are particularly used in \ac{iot} environments, where data comes from various sources like sensors or mobile devices.
However, a central use case for streaming queries is processing on edge-cloud infrastructure, where resources have highly varying capacities in terms of compute, memory, and network.

\textbf{Operator placement for \ac{iot}-scenarios is hard.} 
One major challenge in \ac{iot}-scenarios involving heterogeneous hardware spanning from very simple edge devices to server-grade machines in cloud data centers is \emph{finding} and \emph{reasoning} about operator placement to achieve high performance.
For instance, placing a stream processing operator on a hardware resource located very far from the data source would result in very high network latencies and, hence, overall high end-to-end latency for detecting certain events. 
Likewise, a low-performing edge device with restricted CPU resources will impact throughput if too many computations are executed on it simultaneously. 

\textbf{The initial operator placement is crucial.}
Given the heterogeneity of devices in \ac{iot} scenarios, the initial operator placement is crucial and highly challenging.
In fact, a ``bad'' initial placement can lead to fatal execution behavior, e.g., due to a placement of computationally intensive operators to weak hardware resources.
One substantial consequence of a bad initial placement is high \textit{backpressure} at runtime, where the internal queues of a \ac{dsps} quickly fill up, leading to delays and even query crashes.
Furthermore, an initial good placement is also crucial to avoid expensive operator migrations at runtime, which are especially costly since operators and state needs to be moved. 
Therefore, finding an optimal initial placement is extremely important in these scenarios to avoid data losses noticeable performance drops, or even crashes.
However, finding an initial placement given heterogeneous hardware is particularly difficult without knowing the runtime behavior of a query on that hardware. 

\begin{figure}[t]
    \centering
    \includegraphics[width=0.93\linewidth]{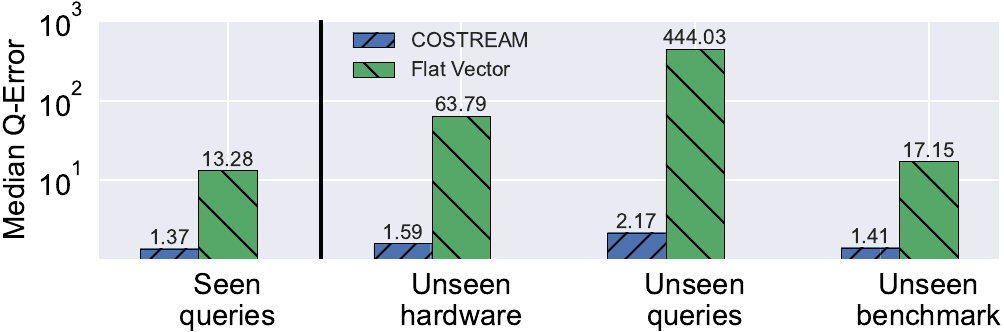}
    \vspace{-1ex}
    \caption{Estimation errors when predicting E2E-latency for queries that are similar to the training data (left) or entirely unseen in terms of underlying hardware and other query properties (right). \costream can precisely predict query execution costs compared to an existing cost model baseline (Flat Vector).}
    \label{fig:motivating_plot}
    \vspace{-5.0ex}
\end{figure}

\begin{figure*}[ht]
    \centering
    \includegraphics[width=1.0\linewidth]{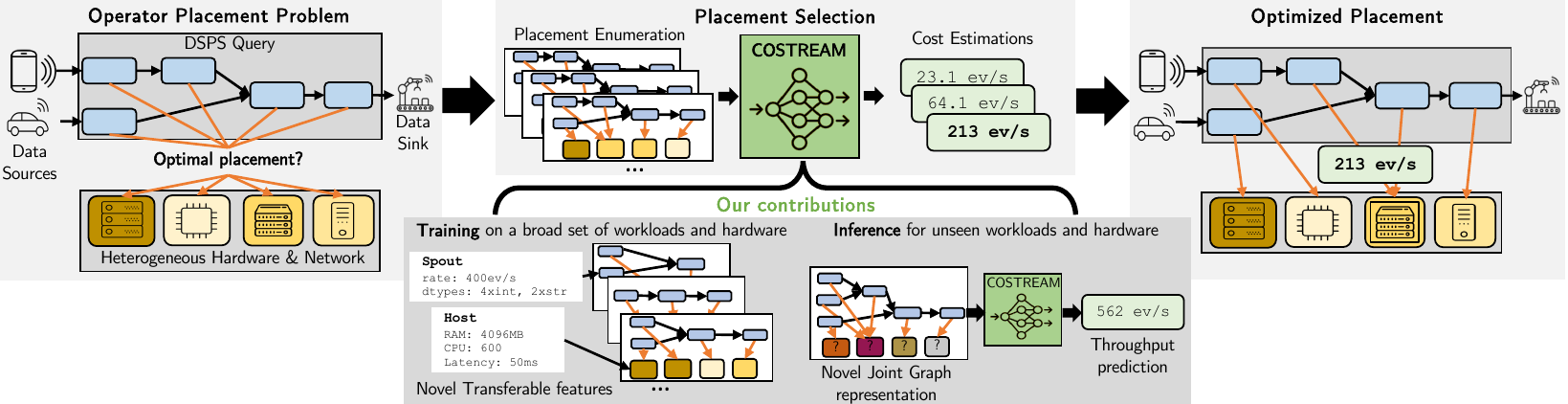}
    \vspace{-4ex}
    \caption{Overview of our approach, which uses a learned cost-model \costream{} for operator placement. \costream{} is trained by a zero-shot approach on a broad set of workloads and hardware and thus can infer costs even for unseen workloads and hardware. }
    \label{fig:high_level_idea}
    \vspace{-4ex}
\end{figure*}

\textbf{Issues of existing placement approaches.} 
Although the operator placement problem has gained significant attention in prior research~\cite{aniello2013, imai2017, chandramouli2011, heinrich2022, pietzuch2006, wang2017, cardellini2016, nardelli2019, luthra2021, liu2023, eskandari2019}, there are notable shortcomings in the existing approaches. A primary limitation lies in the predominant emphasis on \textit{online reconfiguration} during query execution, neglecting the crucial need for initial (offline) placement.
Furthermore, a significant gap exists in addressing hardware and network heterogeneity, particularly crucial in \ac{iot}-scenarios. While some papers acknowledge the presence of heterogeneous hardware~\cite{ni2020, xu2014, luthra2021}, they remain tailored for online reconfiguration, heavily reliant on runtime statistics collected through monitoring, and thus inevitably not usable for initial placement decisions.
Another downside of monitoring approaches is the time they take to adjust the placement to a more optimal one, which in turn causes
non-negligible overheads due to costly operator migrations~\cite{pietzuch2006, aniello2013, eskandari2019, nardelli2019}. 

\textbf{A learned cost model for initial placement.}
In this paper, we present a novel learned cost model \costream{} that can be used to determine the initial placement of operators.
The main idea is that our model predicts the expected performance of a streaming query before running the query, which can be used for optimally placing operators on different hardware resources to maximize query performance.
In contrast to existing approaches for learned operator placement \cite{alnafessah2021, wang2017}, our model does not rely on runtime information and thus enables an initial placement selection.
However, due to missing runtime information, the prediction problem becomes more challenging since performance metrics need to be predicted simply based on characteristics, that are available before execution.

\textbf{Novel model architecture.}
To enable our cost model, we thus developed a novel model architecture based on \ac{gnn}, which represents all available information about query, data, and hardware in one joint graph.
This allows the model to accurately reason about the expected performance of a query when placed on certain hardware resources.
An important property of our model is that it can generalize out-of-the-box to query patterns and hardware resources the model has not seen during training.
As such, the model falls into the category of \textit{zero-shot} cost models \cite{hilprecht2022}.
This property is crucial for preventing the need for constant model retraining whenever new hardware becomes available or when executing previously unseen queries.
We achieve this by carefully selecting transferable features that allow our model to generalize to unseen queries and hardware. 
In \Cref{fig:motivating_plot}, we show the accuracy of our cost model \costream{} (q-error $1$ being a perfect estimate) in comparison to an existing learned approach \cite{ganapathi2009} (Flat Vector) for cost prediction, that does not target the initial placement problem of streaming queries.
In contrast to this baseline, our model predicts query cost highly precise for both \textit{seen} and \textit{unseen} workload and hardware.
While generalizability across unseen \ac{dsps}s is another interesting dimension, we focus on {unseen} workload and hardware.

\textbf{Why cost-based operator placement?}
One could argue about other possible ways for learned operator placement of streaming queries. 
A more direct way, instead of using a cost model, is to apply \emph{end-to-end learning}~\cite{luthra2021, li2018, mao2019} that tries to predict the placement of operators to individual resources directly. 
We, instead, argue for a \emph{cost-based} placement where a cost model is combined with a search heuristic that enumerates different placement options and uses the cost estimates to select the best option.
Unlike end-to-end models, our method grounds decisions in the underlying cost and thus is naturally more transparent (i.e., one can easily debug a placement decision based on the predicted costs).
Moreover, using a cost-based approach paves the road for potential extensions to solve even more complex optimization problems in the future, such as offline operator reordering~\cite{hirzel2014} or selecting optimal parallelization degrees offline ~\cite{agnihotri2023}.  
The design choice of using a cost model for query optimization also finds validation in established query optimizers within database systems~\cite{leis_2015, siddiqui_2020}. 

\vspace{-0.5ex}
\textbf{Contributions of the paper.} 
To summarize, this paper makes the following contributions:
(1) We present \costream{}\footnote{\label{footnote:code}Source code at \url{https://github.com/DataManagementLab/costream-public}; experimental data and trained models at \url{https://osf.io/5ktgv/}}, a learned zero-shot cost model for operator placement on heterogeneous hardware (\Cref{sec:costream}).
(2) We present our selection of transferable features and cost metrics required to reason about the placement on a given hardware resource (\Cref{sec:realization}).
(3) We further show how \costream{} can be used to solve the initial operator placement problem before executing a query and show that our approach can significantly outperform placements using current heuristics targeting the same problem (\Cref{sec:optimizer}).
(4) We developed a novel cost estimation benchmark, which is a corpus of query executions on heterogeneous hardware that can be used by other researchers in the future (\Cref{sec:training_inference}).
(5) Finally, we use this benchmark for an extensive set of experiments to evaluate \costream{} and show its generalization capability on unseen combinations of hardware, network, and query properties (\Cref{sec:evaluation}).

We also want to note that this paper is based on an existing short paper~\cite{heinrich2022} but significantly extends its contributions. 
The short paper only outlined the basic idea of cost-based placement without taking detailed hardware properties into account. 
Different from \cite{heinrich2022}, the cost model in this paper proposes a novel joint operator-resource graph representation, which is needed to support heterogeneous hardware and co-location of operators on resources. 
Moreover, we devised a new learning procedure to better capture the effects of hardware on query cost.
We will explain the details of all these contributions in the remainder of this paper.

\vspace{-1ex}
\section{\costream{} Overview}
\vspace{-1ex}
\label{sec:overview}
The overall approach of our cost-based placement is shown in \Cref{fig:high_level_idea} and explained in the following.  
The main idea is to use a learned cost model as a major building block to find an operator placement on heterogeneous hardware resources. 
A key aspect of our cost model is that it can be used for accurately predicting cost metrics, even for unseen workload and hardware combinations.
In the following, we give an overview of the training and usage of \costream. 

\textbf{Training the zero-shot model.}
Building a model to precisely predict query cost metrics for edge-cloud scenarios is challenging, as these depend on various factors like the characteristics of the data streams, the operators in the query, and the heterogeneous hardware resources.
In this work, we solve this task by presenting several new ideas:

(1) First, we introduce a \emph{novel joint graph representation} as input to our cost model that covers information about the data streams, the operator graph, and the hardware resources, including the data flow as well as the operator placement (cf. \Cref{sec:costream}).
This allows our model to learn query costs (cf. \Cref{subsec:costs}) required for reasoning about operator placement from all these aspects while taking complex non-linear effects between them into account.
For instance, a windowed operator placed on a node with limited memory resources can significantly suffer from state that needs to be spilled to the disk if the window is too large, which largely influences the latency.

(2) As a second idea, we propose a \ac{gnn}-based model architecture that comes with a novel effective learning procedure to predict costs for given operator graphs and their placements on heterogeneous hardware. 
Earlier work exists, which also applies \ac{gnn}s to predict query costs \cite{hilprecht2022, heinrich2022}, but did not take hardware resources and placement into account.
To ideally support hardware and placement information for cost predictions, we developed a novel strategy of neural message passing that we discuss in Section \ref{sec:costream:scheme}

(3) A third idea is to use \textit{transferable features}, which are applied to describe data streams, query operators, and hardware in a generalizable way (cf. \Cref{subsec:transferable_features}). 
Instead of using features that are strictly tied to a given resource (e.g., \emph{hostname}) or a given workload (e.g., \emph{filter literal}), we identify general features (e.g., amount of memory, network speed, event rates) that allow a model to better generalize to unseen workload and hardware configurations, which is important to find a good initial placement.

\textbf{Using the zero-shot model.} 
Once \costream is trained, it can be used for the initial operator placement of a streaming query.
To find an operator placement, instead of an exhaustive enumeration, which would not be possible for complex queries and landscapes with many different hardware resources, this work uses a search heuristic (cf. \Cref{sec:optimizer}) that is designed for typical \ac{iot}-scenarios to enumerate different alternative operator placements for a given query.
However, other enumeration strategies could also be used jointly with \costream.
We show in our evaluation, that our cost estimations are highly accurate and help to solve the operator placement problem. 
Moreover, we see \costream{} as a starting point to enable other cost-based optimizations such as operator re-ordering or selecting the degree of parallelism~\cite{agnihotri2023}.

\section{The \costream  Model} \label{sec:costream}
\vspace{-1ex}

\begin{figure*}
    \centering
    \includegraphics[width=0.9\textwidth]{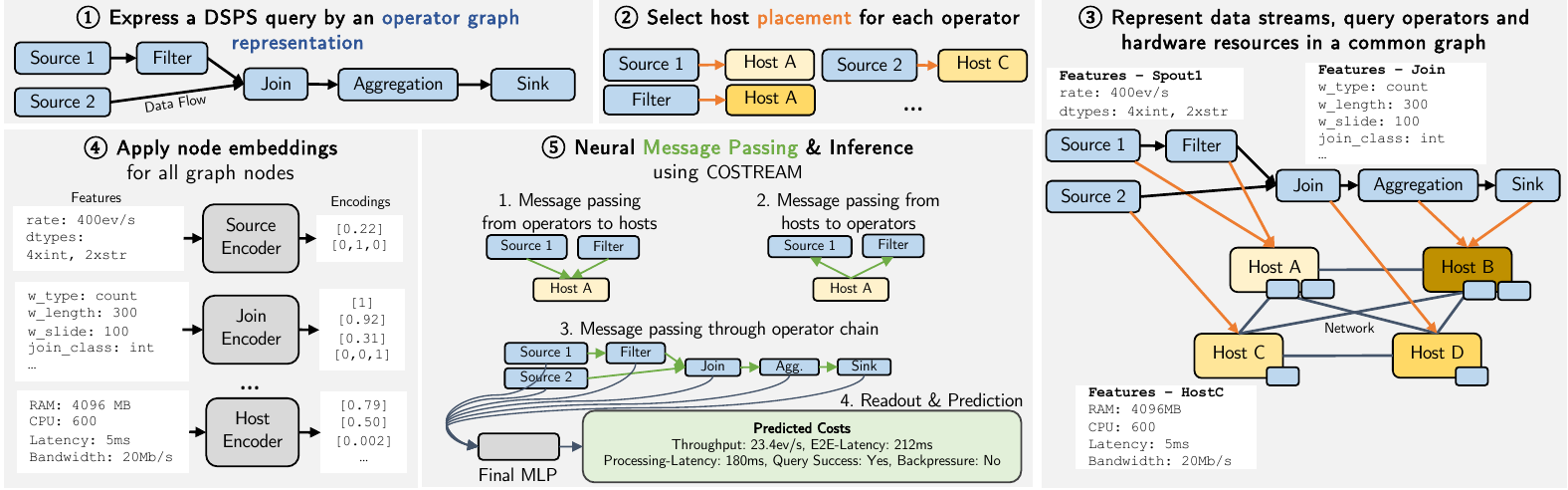}
    \vspace{-1ex}
    \caption{Cost estimation with \costream: \circles{1} At first, a \ac{dsps} query is represented as a \textcolor{CornflowerBlue}{graph of operators} that form a \ac{dag}. \circles{2} For each operator, a corresponding  \textcolor{Goldenrod}{host} is selected, at which it should be located. This is the \textcolor{orange}{operator placement}. \circles{3} The data streams, query operators, and hardware nodes, as well as the data flow and placement are represented as a common learnable graph. \circles{4} Each graph node is described with transferable features that are embedded by node-type specific encoders into hidden states. \circles{5} These states are passed along a \textcolor{LimeGreen}{message passing} scheme through the graph. A separate MLP finally transforms the hidden state into a cost prediction.}
    \label{fig:overall_approach}
    \vspace{-4ex}
\end{figure*}

The approach of \costream{} is illustrated in \Cref{fig:overall_approach}, where we describe how execution costs are predicted for an arbitrary streaming query. 
The question we aim to answer is: ``\textit{What will be the costs of a query given the placement on a specific set of hardware nodes?}'' 
\circles{1} At first, a streaming query is transformed into an operator graph representation, where nodes represent operators and edges represent the data flow. 
This representation is used later to encode the transferable features into a \ac{gnn} that facilitates the learning process. 
\circles{2} In the next step, the mapping of operators (orange edges) on the hardware nodes is selected (yellow nodes) for which the cost predictions are made.  
\circles{3} The overall graph representation comprising the data flow together with the operator mapping on hardware nodes is annotated with its transferable features. 
\circles{4} The graph node features are embedded into vectors called \textit{hidden states} using \textit{encoders} to apply neural message passing \cite{gilmer2017}. 
\circles{5} Finally, neural message passing across multiple directions, i.e., data flow in the operator graph, and bidirectional operator to hardware mapping is performed to infer the initial operator placement costs using several cost metrics.
We explain our novel representation and the learning procedure in the following.

\vspace{-1ex}
\subsection{A joint representation} \label{subsec:representation}
\vspace{-1ex}

As a key contribution of \costream{}, we propose a novel joint representation of data, query, and hardware configurations to predict relevant cost metrics.
In this representation, a \ac{dsps} query is represented as a set of streaming operators ($\omega \in \Omega$) that each operates on one or multiple unbounded input streams ($d \in D$) on a set of computing nodes ($n \in N$) and returns one or multiple output tuples.
Typically, a \texttt{source} operator $\omega_s$ describes the data characteristics of an input stream $d$ into the \ac{dsps}. 
The final operator $\omega_N$, referred to as \texttt{sink}, is responsible for persisting or forwarding the resulting tuples. 
The data flows between the operators $\omega_s \rightarrow \omega_2 \rightarrow \dots \rightarrow \omega_N$ are referred to as \textit{logical data flow} in the following.
In this work, we focus on algebraic streaming operators, namely \texttt{filter} $\omega_\sigma$, \texttt{windowed aggregation} $\omega_\xi$, and \texttt{windowed join} $\omega_{\Join}$ that apply certain computations on the data stream. 
While filter and aggregation operators execute on single incoming data streams, join operators combine incoming tuples from two streams that arrive in a given window. 
Thus, the logical data flow is not always linear but can take the form of a tree. 

In \ac{dsps}, each streaming operator $\omega_i$ is assigned to one compute node $n_j$ (i.e., hardware resource), which is referred to as \textit{operator placement} ($\omega_i \rightarrow n_j$).
In turn, each compute node can execute one or multiple operators.
In \ac{iot}-scenarios, these nodes can be geo-distributed and heterogeneous in their computation capabilities, network speed, etc. 
To model the effects of network communication, we model the network characteristics between pairs of nodes that are used to ship data from one node to another and thus describe the \textit{physical data flow} across the network.

\textbf{GNN-based model.}
Since varying numbers of hosts and operators can occur in given queries and placements, a learning method is required that can deal with these flexible structures. In this work, we propose the use of \ac{gnn}s, which are very well suited for these structures.
In contrast to previous work, \cite{heinrich2022, hilprecht2022} that describes only the operator graph, our idea is to include hardware resources and the location of operators.
We thus represent data sources and sinks, query operators, and hardware in one joint graph.
In particular, we model the query operators $\Omega$ in a \ac{dag}, where each vertex represents an operator $\omega_i$, and the directed edges between these represent the logical data flow as shown in \circles{1}. 
In addition, each hardware instance is represented in this graph as vertex $n_i$. 
For each operator node $\omega_i$, we model the operator placement $\omega_i \rightarrow n_j$ as a mapping from operators to hardware resources as shown in \circles{2}. 
Putting it all together, a joint \ac{dag} results that contains operator and hardware nodes and describes the logical data flow and the operator placement by the directed edges shown in \circles{3}. 
Each operator node $\omega_i$ and each computing node $n_j$ is described by a feature vector $v_i$. 
This procedure leads to a GNN-based query representation used to train \costream.

\vspace{-1ex}
\subsection{Training procedure of \costream{}}
\label{sec:costream:scheme}
\vspace{-1ex}

\costream{} is trained with a \ac{gnn}-architecture using a novel message passing scheme. 
As shown in \circles{4} of \Cref{fig:overall_approach}, we first pass the transferable features from all graph nodes as feature vectors $v_i$ to so-called \textit{encoders}. 
The encoders are multi-layer perceptrons (MLPs) that encode the features into fixed vectors called hidden states $h_v$.
For each node type $T$ (e.g., source, join, host, etc.), we apply a separate encoder $MLP_{T(v)}$ as shown in.
In the next step, we use these initial states as a foundation for the \textit{message passing} (shown in \circles{5}) that is used in \ac{gnn} to learn from node neighborhoods~\cite{gilmer2017}. 
The hidden states of each graph node are updated over multiple iterations by combining the incoming hidden states from its children nodes with the current state. 
For every graph node $v$, all hidden states of previous nodes are summed and concatenated with their hidden state $h_v$ \cite{zaheer2017}. 
This intermediate hidden state is then fed into another node-type specific $MLP'_{T(v)}$ to obtain the updated hidden state $h'_v$ of node $v$. 
We summarize the algorithm of the learning procedure in \Cref{alg:message_passing}.

\begin{figure}[t]
\vspace{-2ex}
\begin{algorithm}[H]
	\scriptsize
    \captionsetup{font={scriptsize}}
	\caption{Learning procedure of \costream}
    \label{alg:message_passing}
    \textbf{Input: } Graph $g$ with operator and hardware nodes $v$ and input features $x(v)$ \\
    \textbf{Output:} Cost prediction $C$
	\begin{algorithmic}[1]
        \For{$v\in$ $g$} \Comment{Compute hidden state per node} \label{alg:line:embedding}
        \State $h_v \leftarrow \mathit{MLP}_{T(v)}(x_v)$
        \EndFor
        \For{order $\in$ (\textsc{ops$\rightarrow$hw}, \textsc{hw$\rightarrow$ops}, \textsc{sources$\rightarrow$ops})}
            \For{$v\in$ order} \Comment{Updates according to message passing order}
            \State $h'_v$ $\leftarrow$ $MLP'_{T(v)}\left(\sum_{u\in\mathit{children(v)}}h'_u + h_v\right)$ 
            \EndFor
        \EndFor
        \State $C\leftarrow MLP_{\mathit{out}}(\sum_{v\in\mathit{g}})h'_v$ \Comment{Estimate costs with a graph readout}
        \State \Return $C$
	\end{algorithmic}
\end{algorithm}
\vspace{-7ex}
\end{figure}

\textbf{Novel message passing scheme.} A key challenge to solve in the training process is to decide on the message-passing scheme in the graph representation, as this is very important for precise cost predictions for the given placement. 
Next, we discuss our message-passing scheme:

\textbf{(1) Operators to hardware (\textsc{ops}$\rightarrow$\textsc{hw}):} 
At first, the hidden states are passed from all operator nodes $\omega_i \in \Omega$ to their corresponding host nodes $n_j \in N$. 
The intuition of this step is to inform the host nodes about the computational requirements of the operators executed on the hosts.
Note that in the case of co-location, multiple messages coming from different operator nodes are passed to the host nodes.

\textbf{(2) Hardware to operators (\textsc{hw}$ \rightarrow$ \textsc{ops}):} 
Then, we pass the combined hidden states back to the initial operator nodes. 
The intuition of this step is to inform the operator nodes about the host nodes that they are placed on.

\textbf{(3) Data sources to operators (\textsc{sources} $\rightarrow$ \textsc{ops}):} 
Afterward, we apply message passing following the data flow through the operator chain until we arrive at the sink $\omega_N$.
This allows the propagation of characteristics of data sources (e.g., event rates) through the operator graph and merged with the operator and hardware information. 

\textbf{(4) Final readout}: 
After the message passing, the hidden states from all nodes are read out and summed up. The resulting state is then passed to a final $MLP_{out}$ that predicts the overall query costs $C$.

\section{Realization of \costream{}} \label{sec:realization}
\vspace{-1ex}
For making initial placement decisions with \costream{}, we instantiate and train separate \ac{gnn} models to predict different relevant cost metrics.
The cost metrics required for placement decisions are explained in \Cref{subsec:costs}.
Afterward, we describe the selection of transferable features in \Cref{subsec:transferable_features} to enable the prediction of these metrics.

\subsection{Cost metrics and model implementation}
\label{subsec:costs}
\vspace{-1ex}

We identify and choose \textit{five} different cost metrics $C = (T, L_p, L_e, R_O, S)$, that together describe the performance of an initial placement to be predicted by our cost model.
Besides common metrics like throughput $T$ and two kinds of latencies (i.e., processing $L_p$ and end-to-end latency $L_e$), widely used in DSPS \cite{wang2017, karimov2018}, we propose to use backpressure occurrence $R_O$ and the execution success $S$ (both binary) of placement as additional metrics.

\textbf{Discussion:} 
We argue that \textit{all} of the presented metrics are indispensable representatives of query costs to provide high-quality decisions of \costream{}. 
First, avoiding backpressure occurrence $R_O$ and enabling execution success $S$ for a query are instrumental for a ``good'' initial placement. 
To enable high accuracy for both metrics, we model them as classification tasks which are simpler to solve than regression tasks that are needed to predict latency and throughput.
However, as discussed later, the models for all metrics share the same \ac{gnn}-based architecture, and only the final MLP is different depending on the predicted metric.


\textbf{Metrics:} 
In the following, we briefly present the definitions of these cost metrics that our model needs to learn to understand the execution behavior of a \ac{dsps} query for enabling initial placement decisions.

\begin{definition}
\textbf{Throughput (T)}: For the execution of a given query, we define $T$ as the number of output tuples that arrive at the sink per time unit.
\end{definition}

We define the \textit{processing-} and the \textit{end-to-end-latency}.
While the former describes \textit{just} computation and networking transfer latencies within the query execution, the latter includes potential waiting times in a preceding message broker~\cite{akidau2015}.

\begin{definition}
\textbf{Processing latency (L\textsubscript{p}):} For each output tuple $d_O$, $L_p$ is the interval between the time at which the oldest input tuple $d_I$ involved in producing the output tuple $d_O$ is \textit{ingested} in the query and the time that $d_O$ arrives at the sink.
\end{definition}
\begin{definition}
\textbf{End-to-end latency (L\textsubscript{e}):} For each output tuple $d_O$, $L_e$ is the interval between the time at which the oldest input event tuple $d_I$ involved in producing the output tuple $d_O$ is \textit{generated} at the event broker and the time that $d_O$ arrives at the sink.
\end{definition}

If a sub-optimal operator placement is selected and resources are over-utilized, \textit{backpressure} may occur.
In that case, incoming tuples are queued up, leading to a prolonged end-to-end latency \cite{karimov2018,chen2017}. 
Since such cases should be avoided, we introduce a new metric that our cost model predicts:

\begin{definition}
\textbf{Backpressure occurrence (R\textsubscript{O})}: The backpressure rate $R$ is the number of tuples per time unit that are queued up in the message broker of a \ac{dsps} system in case of backpressure. 
If $N$ multiple data streams $d_1$, $d_2, \ldots$ are backpressured, then $R$ is the sum of all single backpressure rates: $R = \sum_{i=1}^N B_i(d_i)$. 
Here, each $B_i(d_i)$ is given as the difference between the arrival and the processing rate for that stream.
If backpressure occurs during the query execution, i.e., $R \geq 0$, we define the backpressure occurrence $R_O$ with $R_O = 0$, else $R_O=1$.
\end{definition}

A \ac{dsps} query execution might be unsuccessful, which can happen due to two reasons: 
(1) Garbage Collection\footnote{Java is the main programming language of all major \ac{dsps}~\cite{carbone2015, toshnival2014, kulkarni2015, akidau2015}.} especially happens when placing memory-intensive operators to low-performing hardware nodes and might lead to application pauses and even crashes. 
(2) Due to logical conditions (low selectivity, short windows), no tuple arrives at the sink during the execution.
We define a binary metric for query success $S$:
\begin{definition}
\textbf{Query success (S)}: The query success $S$ of a \ac{dsps} is $S=0$, if no tuple arrives the sink $\omega_n$ during the execution time given the placement. Else, $S=1$.
\end{definition}

\textbf{Model Implementation:} We train separate \ac{gnn}s for the previously defined cost metrics in $C = (T, L_e, L_p, R_O, S)$.
For encoding query, data streams, and hardware, we used the joint operator-resource graph as discussed in \Cref{sec:costream}.
To predict the metrics $T$, $L_p$, and $L_r$, we trained separate regression models based on the encoding resulting from the joint operator-resource graph.
As our cost metrics for regression tasks have a very large value range, we found the \textit{Mean Squared Logarithmic Error} (MSLE) as an optimal loss function. 
It is defined as: $L(y, \hat{y}) = \frac{1}{N} \sum_{i=0}^N ((\text{log}_e(1 + y_i)) - (\text{log}_e(1 + \hat{y}_i))^2$. 
For backpressure, we predict $R_O$ by a binary classification model instead. 
The reason is that $R$ has an enormous value range while $R=0$ in cases of no-backpressure, making it hard to train a precise regression model.
Similarly, we trained a binary classification model to predict the query success $S$.

Furthermore, a second important aspect of the model implementation is that we apply the idea of \textit{ensemble learning} to improve the certainty of the predictions.
Instead of relying on a single model per metric for prediction, we use multiple separately trained cost models and combine their predictions for the placement decision.
For each model instance, we varied the random initialization seed of each model to find different local minimums in the parameter space to obtain different predictions for the same query.
At inference time, we thus apply a majority vote over the binary predictions (for $S$ and $R_O$) and do a mean computation for all regression models.

\vspace{-1ex}
\subsection{Transferable features for cost prediction} \label{subsec:transferable_features}
\vspace{-1ex}
An important part of our cost model design is the selection of meaningful features, that have to meet two requirements:

(1) The features have to enable a prediction of the costs for an initial placement by describing the most important properties of an execution; i.e. query complexity, workload, and hardware resources.
Moreover, the features have to be determined \textit{before} the execution so that they allow prediction of the initial placement. 
We explicitly set and enumerate different operator, hardware, and data characteristics on the underlying \ac{dsps} to acquire training data on these features. As such, there is no overhead in obtaining these features.

(2) \costream has to reliably predict costs of initial placements for queries and hardware resources that differ from the training data and are thus \textit{unknown}.
Precisely, the model needs to \textit{extrapolate} (i.e. beyond training data range) and \textit{interpolate} (i.e. within the training data range, but differing).
For instance, extrapolation is required for a query running on weaker hardware resources than those previously used in the training.
We thus propose \textit{transferable} features that enable generalizability and meaningful inter- and extrapolation.
We present a complete list of those features in \Cref{tab:features} which are divided into the categories of operator-related, data-related, and hardware-related as follows:




\begin{table}
    \centering
    \tiny
    \begin{tabular}{p{0.6cm}p{0.65cm}p{2.2cm}l}
        \textbf{Node} & \textbf{Category} & \textbf{Feature} & \textbf{Description} \\ \hline
        \multirow{2}{*}{all}
         & data & \texttt{tuple width in} & Averaged incoming tuple width \\
         & data & \texttt{tuple width out} & Outgoing tuple width \\ \hline
        \multirow{2}{*}{source} & data & \texttt{input event rate} & Event rate emitted by the source \\
         & data & \texttt{tuple data type} & Data type for each value in tuple \\ \hline
        \multirow{3}{*}{filter} & operator & \texttt{filter function} & Comparison function  \\
         & operator & \texttt{literal data type} & Data type of comparison literal \\
         & data & \texttt{selectivity} & see \Cref{def:filter_sel} \\ \hline
        \multirow{2}{*}{join} & operator & \texttt{join-key data type} & Data type of the join key \\
         & data & \texttt{selectivity} & see \Cref{def:join_sel} \\ \hline
        \multirow{3}{*}{agg.} & operator & \texttt{agg. function} & Aggregation function \\
         & operator & \texttt{group-by data type} & Data type of group-by attribute \\
         & operator & \texttt{agg. data type} & Data type of each value to aggregate \\
         & data & \texttt{selectivity} & see \Cref{def:agg_sel} \\ \hline
        \multirow{4}{*}{window} & operator & \texttt{window type} & Shifting strategy (sliding/tumbling) \\
         & operator & \texttt{window policy} & Counting mode (count/time-based) \\
         & operator & \texttt{window size} & Size of the window \\
         & operator & \texttt{slide size} & Size of the sliding interval \\ \hline
         \multirow{4}{*}{hardware} & hardware & \texttt{cpu} & Available CPU resources in \% \\
         & hardware & \texttt{ram} & Available RAM resources in MB \\
         & hardware & \texttt{network-latency} & Outgoing latency of the host in ms\\
         & hardware & \texttt{network-bandwidth} & Outgoing bandwidth of the host in Mbit/s\\ \hline
    \end{tabular}
    \vspace{-2ex}
    \caption{Overview of transferable features. These features apply to any streaming workload and hardware configuration. \costream{} learns from these features to predict query execution costs. They can be divided into operator-, data-, and hardware-related features}
    \label{tab:features}
    \vspace{-9ex}
\end{table}

\textbf{Operator-related features}. 
Operator-related features enable the model to take the query complexity into account when making performance predictions.
As such, these features must be sufficient for the model to implicitly derive aspects such as the computational and memory complexity of an operator, which is important to make performance predictions for different hardware resources.
Intuitively, the model can derive the memory and computational requirements of an operator from operator-related features. 
Then the model can make predictions of the operator performance placed on certain hardware given its resources, (e.g., amount of memory, speed, and number of cores), which we model by the hardware-related features (see below).
To indicate the base complexity of operators, we use the \texttt{operator type} as a main feature (e.g., filter or a join).
Moreover, each operator comes with further operator-specific features.
For example, to infer relevant memory requirements for stateful operators (i.e., windowed operations), we use information such as the \texttt{window length} and its \texttt{window type}. 
Another example of operator-specific features is the complexity of filter predicates.
For this, we model the predicate structure (i.e., how many filters) as well as data types of filter constants.
To train our model with these features, we create query plans and deliberately set the values of these features to cover a wide spectrum of queries (e.g., different window sizes).

\textbf{Data-related features}.
Describing the query operators alone is not sufficient, as query costs depend on the data characteristics as well.
For example, the execution frequency of a count-based window depends on its tuple arrival rate.
As such, we model the tuple ingestion rate at the data sources as one of the main features along with the data type for all attributes in a tuple.
Moreover, we also need to be able to express this rate not only for the sources but also derive the tuple arrival rates for operators that operate on the output of other operators (i.e., further downstream in a plan).
For this, we annotate the \texttt{selectivity} to each operator. 
While \texttt{tuple width} and expected \texttt{event rate} at the source are given for cost prediction, the selectivities need to be estimated, since they are not available before the runtime of a query. 
For this, we first define the selectivity for all operators we support; i.e., \texttt{filter}, \texttt{windowed join} and \texttt{windowed aggregation} according to our previous work~\cite{heinrich2022}:

\begin{definition}
    \label{def:filter_sel}
    \textbf{Filter selectivity ($\bm{\text{sel}({\omega_\sigma}))}$}:
    The selectivity $sel({\omega_\sigma})$ of a filter operator $\omega_\sigma$ is the ratio of the number of outgoing to incoming tuples in the input stream $D$:
    \vspace{-1ex}
    \begin{equation*}
        sel ({\omega_\sigma}) = \frac{|f_{\omega_\sigma}(D)|}{|D|}, \quad  \text{with   } 0 \leq sel ({\omega_\sigma})  \leq 1.
    \end{equation*}
    \vspace{-2ex}
\end{definition}

\begin{definition}
    \label{def:join_sel}
    \textbf{Join selectivity ($\bm{\text{sel}({\omega_{\Join}}))}$}:
    The selectivity $sel(\omega_{\Join})$ of a \emph{windowed join} operator that considers tuples from windows $W_{d_1}$ and $W_{d_2}$ over two input streams $d_1$ and $d_2$ is the ratio of qualifying join partners to the cartesian product for all tuples in the input windows:
    \vspace{-1ex}
    \begin{equation*}
        \label{eq:join_sel}
        sel (\omega_{\Join}) = \frac{|W_{d_1} \Join W_{d_2}|} {|W_{d_1} |\times |W_{d_2} |}, \quad  \text{with   }  0 \leq sel (\omega_{\Join})  \leq 1.
    \end{equation*}
    \vspace{-2ex}
\end{definition}

\begin{definition}
    \label{def:agg_sel}
    \textbf{Aggregation selectivity ($\bm{\text{sel}({\omega_\xi}))}$}:
    The selectivity $sel(\omega_\xi)$ of a \emph{windowed aggregation} operator that considers tuples in a window $W$ from an input stream $D$ is the ratio of distinct group-by values in the window over the window length:
    \vspace{-1ex}
    \begin{equation*}
        \label{eq:agg_sel}
        sel (\omega_\xi) = \frac{|\text{\emph{group-by }}(W_D)|} {|W_D|}, \quad  \text{with   }  0 \leq sel (\omega_\xi)  \leq 1.
    \end{equation*}
    \vspace{-2ex}
\end{definition}

The question arises of how to obtain selectivities, as these are unknown before the query execution. 
Since we aim to predict the cost for initial placement, we rely on existing estimation techniques for selectivity~\cite{dutt2019}, which require a representative sample of the processed data streams.

\textbf{Hardware-related features.} 
Finally, the placement costs that we aim to predict are not only determined by the query complexity and the workload but also by the underlying hardware.
For instance, the costs of a windowed operator are increased if the underlying available RAM is too small as explained below.
As \costream{} supports heterogeneous \textit{unseen} hardware, it needs to be encoded in a transferable way, as resources can differ from those seen during training.
Describing computing and networking resources is a non-trivial task, as these are complex in their behavior and inner architecture.
Therefore, we empirically analyzed the behavior of distinct parameters on our cost metrics and selected four metrics to describe heterogeneous resources.
Note that these features are typically readily available from the hardware itself or can be easily obtained, e.g., from cloud providers.

\textbf{(1) Compute resources:} We encode the available {CPU resources} that are assigned to an operator using a relative metric as a feature; i.e., 200\% of CPU resources refers to a machine having double the compute resources (e.g., 2 cores or 1 core with doubled speed) compared to a single reference core.
Such relative metrics for CPU resources are often used by cloud providers as well to describe the available compute resources in a machine.

\textbf{(2) Memory resources:} 
As another feature, we use the {amount of memory} (RAM) in a machine which has a strong effect on the performance of \ac{dsps}, in particular, if queries with state are executed. 
If the available amount of RAM is too small, this will affect the overall query costs due to swapping and garbage collection. 
We decided not to model RAM speed (i.e., bandwidth and latency) since this only minimally influences the performance of many streaming engines. 
However, such features could easily be added to our model.

\textbf{(3) Network resources:} We further model the maximum network bandwidth of each machine, as this can be a limiting factor for very data-intensive workloads.
Especially in \ac{iot}-settings, the network bandwidth from the edge to the cloud might be much smaller than between a cloud server.
Thus, we encode both {bandwidth} and {latency} as relevant features to decide on an initial operator placement.
\begin{figure}
    \centering
    \includegraphics[width=0.9\columnwidth]{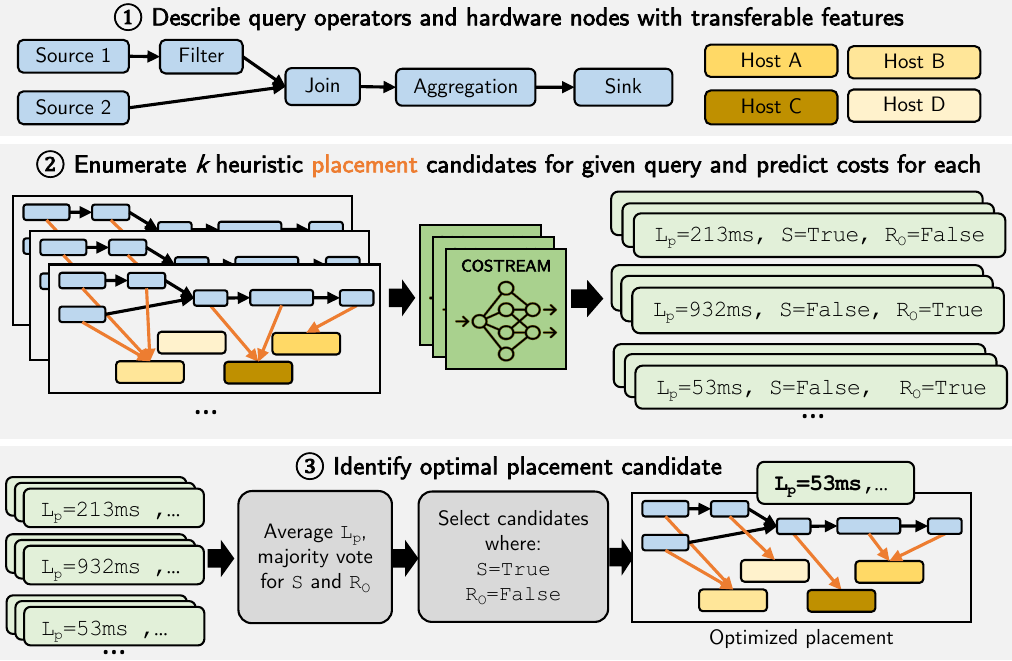}
    \vspace{-1ex}
    \caption{Optimizer model. 
    \circles{1} The operator and hardware nodes are described using transferable features. 
    \circles{2} $k$ placement candidates are generated by randomly distributing the operators to the hardware nodes and using parallel \costream{} instances to predict the query execution costs.
    \circles{3} We average predictions of the target metric ($L_p$ in this example), filter out candidates that are predicted as being backpressured or unsuccessful, and choose the one with the lowest cost, which is the resulting placement.}
    \label{fig:optimizer_model}
    \vspace{-4.5ex}
\end{figure}

\section{Placement Selection with \costream{}}
\label{sec:optimizer}
\vspace{-1ex}

In this section, we explain how cost estimates can be applied to solve the initial operator placement problem.
To solve this problem, we use our \costream model to estimate query costs for a given placement.
By enumerating and estimating different placement candidates, the corresponding estimates can then be compared to identify an optimal one.
Notice that all of our presented cost metrics in \Cref{subsec:costs} are crucial for reasoning about the performance of a given placement candidate.
In our approach, \textit{one} of these metrics is used as a target (e.g., minimizing $L_p$), chosen by the user according to the overall optimization goal.
As the query execution might fail or be under backpressure, predicting $S$ and $R_O$ is additionally required as a sanity check before deciding on placement.
In the following, we outline the procedure for finding a placement.

\textbf{Placement procedure.}
\Cref{fig:optimizer_model} shows how we solve the initial operator placement problem with \costream.
\circles{1} We describe a given query consisting of operator and hardware nodes using transferable features, as explained before. 
\circles{2} We then create a set of placement candidates for the given query operators. 
In this procedure, we selected a heuristic enumeration strategy based on \cite{chaudhary2020}, aiming to represent realistic placements in \ac{iot}-scenarios on certain hardware resources.
This strategy is explained below in more detail.
However, in general, any enumeration strategy can be combined with our cost model.
Afterward, predictions for all placement candidates are obtained with \costream. 
\circles{3} We now identify the optimal placement candidate. 
First, all candidates are filtered out that are either predicted as being not successful or showing backpressure. 
Since we use an ensemble of models, as discussed before, we do a majority vote over the binary predictions (for $S$ and $R_O$) to predict whether a placement results in a successful execution and has no backpressure.
Afterward, for the remaining placements, we select a placement based on the predicted target metric (i.e., one that maximizes $T$ or minimizes $L_p$). 

\begin{figure}
    \centering
    \includegraphics[width=\columnwidth]{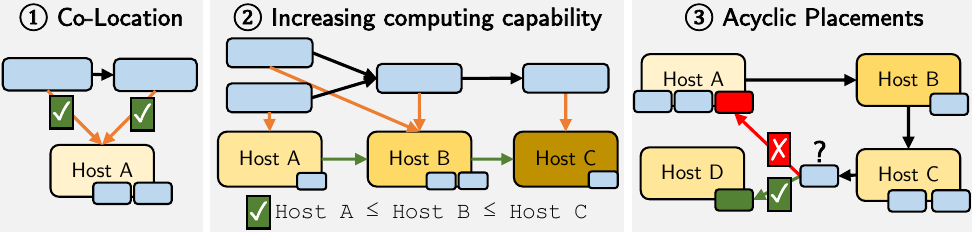}
    \vspace{-4ex}
    \caption{Rules for placement enumeration in our benchmark. \circles{1} Operator co-location, \circles{2} increasing computing capability along the physical data flow, \circles{3} acyclic placements.}
    \label{fig:heuristics}
    \vspace{-5ex}
\end{figure}

\textbf{Heuristic search strategy.} 
For enumerating placement candidates, one challenge is to explore the vast space of possible initial placements.
This work focuses on \ac{iot}-scenarios as a critical application for \ac{dsps}, and thus, we apply an appropriate search strategy.
In these scenarios, data typically flows from sensors to more powerful resources, e.g., from weaker nodes at the edge to more powerful nodes in the cloud.
To reflect such placements, we adapt heuristics for the enumeration procedure (based on~\cite{chaudhary2020}) for the initial placement problem as shown in \Cref{fig:heuristics} and explained as follows:

\textbf{\circles{1} Operator co-location:} In edge-cloud scenarios, the same hardware resources can be used for multiple queries or multiple operators of the same query. 
As such, we allow co-location of multiple operators $(\omega_i, \omega_j, \dots) \rightarrow n_k$ on the same host as this reflects a typical optimization approach to reduce network latencies.

\textbf{\circles{2} Increasing computing capability:} We assume that data is always passed from weaker to stronger instances ($n_i \rightarrow n_j)$, which is a realistic scenario.
For instance, in \ac{iot}-scenarios, data will be streamed from sensors and edge devices to stronger workstations or cloud servers. 
We apply this constraint by classifying hardware into three different bins. 
For each operator placement $\omega_i \rightarrow n_j$, we ensure that all subsequent placements along the data flow are assigned to hardware nodes $n_k, n_l, \dots$ that have the same or a stronger instance category than $n_j$. 
These bins are intersected in their feature range to emulate realistic transitions.

\textbf{\circles{3} Acyclic placements:} As mentioned before, in many real-world scenarios data flows in one direction.
For placements, this means that we do \textit{not} send data back and forth between two nodes; i.e., if data once has passed a computing host $n_i$, it must not be sent back to a host $n_j$, that has previously visited.
We exclude these placements, as they incur network utilization overhead and thus are inefficient and unlikely to be chosen.
\vspace{-1ex}
\section{A new cost estimation benchmark} \label{sec:training_inference}
\vspace{-1.5ex}
For the cost estimation of initial placements, we created a new benchmark of 43,281 query traces since no such benchmark exists for \ac{dsps}.
The benchmark covers a high variety of different queries with various patterns (from simple to complex queries), a variety of hardware resources, and different operator placements, as well as the resulting cost metrics for this placement.
We use our benchmark along with public existing benchmarks to evaluate \costream{} in \Cref{sec:evaluation}.
We plan to release this benchmark to the community, which we believe will be an interesting resource to enable future research on learned cost-based optimization.

\textbf{Hardware variety in benchmark.}
As hardware heterogeneity is important for our benchmark, we applied \textit{hardware virtualization} on physical machines to mimic heterogeneous instances flexibly and efficiently.
Such physical hardware virtualization is a typical mechanism used by cloud providers to provide machines that generate negligible virtualization overhead.
Precisely, for \costream{}, we used bare-metal instances and applied Linux \texttt{cgroups} (developed by Google) to limit the available resources for \ac{dsps} operators.
Physical hardware virtualization allows to have (virtualized) compute nodes with different CPU, RAM, and network capacities while achieving resource isolation between computing nodes at the same time.
The configured resources directly translate to the hardware-related features required by \costream.
Moreover, for training, we also need to select different placements for operators and then run the query plan to collect training labels for throughput and latency.
For this, we use \texttt{cgroup} to define different machine types and \textit{pin} operators to the corresponding \texttt{cgroup}.
Network bandwidth and latencies are
also defined for the machine types by using \texttt{netem}.

\begin{figure}[t]
\centering
\includegraphics[width=0.8\linewidth]{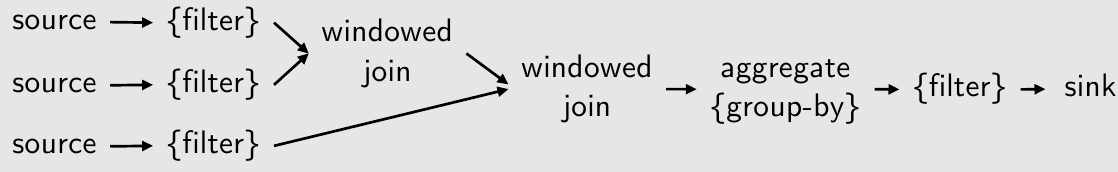}
\vspace{-1ex}
\caption{Example 3-way-join query template for training data generation. \emph{Filter} operators and \emph{group-by} are optional.}
\label{fig:queries}
\vspace{-4ex}
\end{figure}

\textbf{Query and data of benchmark.}
To collect a representative query workload for learning, we emphasized the generation of queries with standard streaming operators like filters, window-aggregates, and window-joins.
Thus, our query workload\footref{footnote:code} includes a nearly equal distribution of linear filter queries, 2-way-, and 3-way joins (35\%, 34\%, 31\%), which we exemplify in \Cref{fig:queries}.
For each of these queries, we randomly apply common streaming operators like windowed aggregation, filters, joins, and group-by with random properties, like window lengths or window types (count- or time-based) according to the training data range \Cref{tab:features}.
We also include different numbers of filter predicates and aggregates to increase the complexity of the queries. 
In our dataset, 35\% of the queries have 1, 34 \% have 2, 24\% have 3, 6\% have 4 filters, and in half of the queries, we applied an aggregation.
For each data stream in a query, we randomly choose a tuple width and an event rate to simulate different workloads.

\section{Experimental evaluation} \label{sec:evaluation}
\vspace{-1ex}

This section reports the experimental evaluation. 
We present the following questions to assess the accuracy and efficiency of \costream for the initial operator placement problem:

\begin{itemize}[leftmargin=*]
    \item \textbf{Exp 1. -- General prediction accuracy:} How accurately does \costream{} predict in general for different hardware, data, and query characteristics?
    \item \textbf{Exp 2. -- Placement Optimization:} What is the performance of initial placements when using \costream?
    \item \textbf{Exp 3. -- Generalization for hardware (interpolation):} How precisely does the model predict costs for query, data, and hardware characteristics that are unseen but within the training range?
    \item \textbf{Exp 4. -- Generalization for hardware (extrapolation):} How accurate is the model for hardware and network resources that are out of the training range?
    \item \textbf{Exp 5. -- Generalization to unseen query patterns:} How accurately does the model predict costs for queries that are unseen in their structure?
    \item \textbf{Exp 6. -- Generalization to unseen benchmarks:} How does the model predict for unseen public benchmarks (i.e., generalization along all dimensions)?
    \item \textbf{Exp 7. -- Ablation studies:} How do major design decisions of \costream{} affect the prediction accuracy?
\end{itemize}

\begin{table}
    \tiny
    \centering
    \begin{tabular}{ll}
        \textbf{Feature} & \textbf{Training data range} \\ \hline
        \texttt{cpu} & {[}50, 100, 200, 300 400, 500, 600, 700, 800{] \% of a core} \\
        \texttt{ram} &  {[}1000, 2000, 4000, 8000, 16000, 24000, 32000{]} MB \\
        \texttt{network bandwidth} & {[}25, 50, 100, 200, 400, 800, 1600, 3200, 6400, 10000{]} MBits\\
        \texttt{network latency} & {[}1, 2, 5, 10, 20, 40, 80, 160{]} ms \\ \hline
        \texttt{input event rate (linear)} & {[}100, 200, 400, 800, 1600, 3200, 6400, 12800, 25600{]} ev/s\\
        \texttt{input event rate (two-way)} & {[}50, 100, 250, 500, 750, 1000, 1250, 1500, 1750, 2000{]} ev/s\\
        \texttt{input event rate (three-way)} & {[}20, 50, 100, 200, 300, 400, 500, 600, 700, 800, 900, 1000{]} ev/s\\
        \texttt{tuple data type} & {[}3...10{]} $\times$  [\texttt{int, string, double}]  \\ \hline
        \texttt{filter function} & \textless{},\textgreater{},\textless{}=, \textgreater{}=, !=, \texttt{startswith, endswith} \\
        \texttt{literal data type} & \texttt{int, string, double} \\ \hline
        \texttt{window type} & sliding, tumbling \\
        \texttt{window policy} & count-based, time-based \\
        \texttt{window size (count)} & {[}5, 10, 20, 40, 80, 160, 320, 640{]} tuples \\
        \texttt{window size (time)} & {[}0.25, 0.5, 1, 2, 4, 8, 16{]} sec \\
        \texttt{slide size} & {[}0.3 ... 0.7{]} $\times$ window length \\ \hline
        \texttt{join-key data type} &\texttt{int, string, double} \\ \hline
        \texttt{agg. function} & \texttt{min, max, mean, avg} \\
        \texttt{group-by data type} & \texttt{int, string, double, none} \\
        \hline
    \end{tabular}
    \vspace{-1ex}
    \caption{Feature range used by the synthetic training dataset}
    \label{tab:training_data}
    \vspace{-3ex}
\end{table}

\textbf{Evaluation strategy.}   
To estimate the accuracy of our cost model, we use the q-error $q(c, \hat{c})$ for the regression metrics (latencies and throughput).
It describes the relative deviation of a real cost value $c$ and its prediction $\hat{c}$ (i.e., a q-error of $2$ states that cost estimates are factor $2$ off), where $1$ is a perfect estimate.
It is defined as: $q(c, \hat{c}) = max \left(\frac{c}{\hat{c}}, \frac{\hat{c}}{c} \right)$, with $q \geq 1$. 
We report the median (Q50) and 95th percentile (Q95) of the q-error.
For the binary cost metrics ($R_O$ and $S$), we report the accuracy as a percentage of correctly classified queries. 
We split our dataset into a \textit{training}, \textit{validation}, and \textit{test set}, (80\%, 10\%, 10\%) where the latter is only used for the final evaluation. 
For the classification tasks, we balanced the number of test set queries by their binary label to fairly report the prediction ability for both classes.

As \ac{dsps}, we used Apache Storm v2.4.0 \cite{toshnival2014}, and as a data producer we used Apache Kafka~\cite{kreps2011}.
With that setup, we executed the benchmark queries, collected query costs and the DCs, and used them to train \costream{}.
As \ac{dsps} queries are naturally unbounded, we stopped the execution after 4 minutes, and collected labels and DCs from the worker nodes afterward, leading to a total query execution time per query of 5 minutes.

We empirically determined the stability of the query costs for execution time at more than 3 minutes for two practical reasons: (1) The query must run at least long enough to contain several windows to actually achieve output tuples. (2) This time is required by the Kafka Producer to set the desired throughput.
The feature range used for training is shown in~\Cref{tab:training_data}.

\begin{figure*}
\begin{minipage}[t]{\columnwidth}
\vspace{-\topskip}
\centering
    \centering
    \includegraphics[width=\columnwidth]{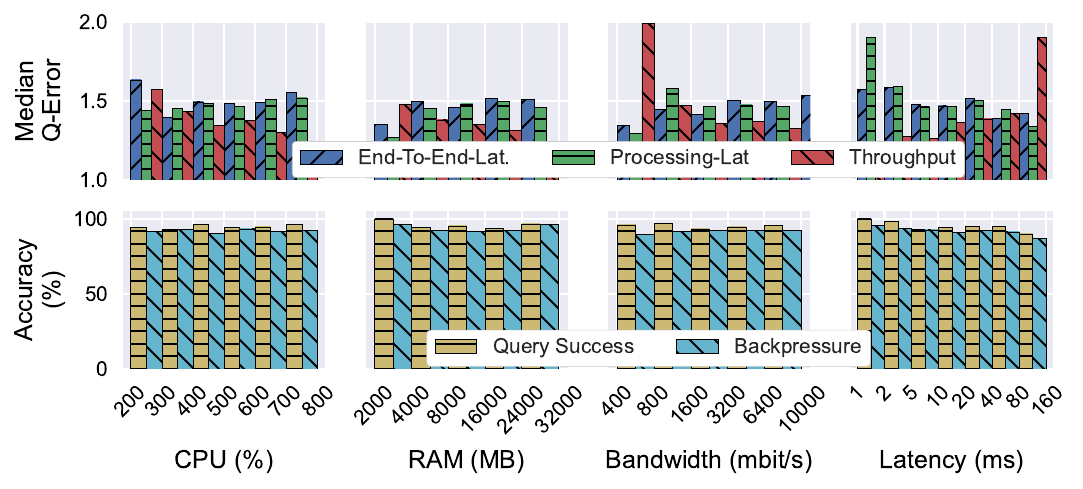}
   \vspace{-4ex}
    \caption{[Exp 1] Prediction results (q-error and accuracy) over hardware and network range. For different average CPU, RAM, bandwidth, and latency ranges \costream{} can precisely predict costs for queries that are executed on heterogeneous resources.}
    \label{fig:qerror_over_hardware}
    \vspace{-3ex}
\end{minipage} \hfill
\begin{minipage}[t]{\columnwidth}
\vspace{-\topskip}
\centering
    \centering
    \includegraphics[width=\columnwidth]{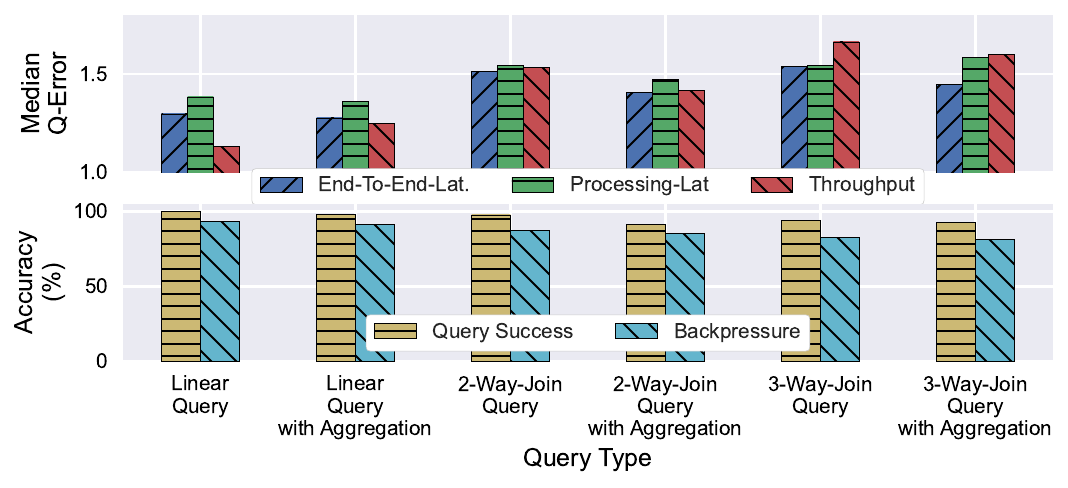}
     \vspace{-5ex}
    \caption{[Exp 1] Prediction results (q-error and accuracy) over the query types. \costream{} can predict costs for all the query types precisely. Q-error increases with the complexity of the query type increases as the overall cost estimation task becomes harder.}
    \label{fig:qerror_over_query_type}
    \vspace{-10ex}
\end{minipage}
\end{figure*}

\textbf{Baselines.}
We compare \costream to a baseline approach for cost estimation~\cite{ganapathi2009} from DBMS.
Since no other cost model for streaming operator placement exists, we extended this model toward streaming queries and placement information.
The model uses a flat vector to represent features that are comparable to \costream, such as input event rates, and query information (e.g., amount of filters). 
Because of the missing structural encoding of features, e.g., related to hardware; not all features can be represented in the flat vector. 
But as shown later, this information is crucial for placement decisions and verifies our model design.
This baseline approach is used to compute a representation (i.e., a vector), on which classification and regression models are trained using \cite{lightgbm} to predict placement metrics $(T, L_p, L_e, R_O, S)$.
Furthermore, we compare against the placement heuristics~\cite{chaudhary2020} and online scheduling approach~\cite{aniello2013} to show speed-ups using our cost estimation for initial placement. 

\textbf{Setup \& implementation.} The training data collection was conducted using CloudLab~\cite{duplyakin2019}. 
To execute training queries based on our novel cost estimation benchmark, we used $60$ available \texttt{m400} machines grouped in 10 clusters.
To provide highly heterogeneous resources for the placements (cf. \Cref{sec:training_inference}), we used Linux \texttt{cgroups} to configure small to large machines using container-like limits on resource usage. 
To model network-wide constraints, we used \texttt{tc-netem}.
\vspace{-1ex}
\subsection{Exp 1: Prediction accuracy} \label{subsec:general_pred_accuracy}
\vspace{-1ex}

\begin{table}
\centering
\scalebox{0.71}{
    \begin{tabular}{lcccc}
        & \multicolumn{2}{c}{\textbf{\costream}} & \multicolumn{2}{c}{\textbf{\textsc{Flat Vector}}} \\
        \hline
        \textbf{Metric} & \textbf{Q50} & Q95 & \textbf{Q50} & Q95\\ \hline
        Throughput & \textbf{1.33} & 5.60 & 9.92 & 590.34\\
        E2E-latency & \textbf{1.37} & 13.28 & 24.96 & 827.59\\
        Processing latency & \textbf{1.46} & 13.90 & 22.87 & 458.14 \\ \hline
        Backpressure & \multicolumn{2}{c}{\textbf{87.89\%}} &  \multicolumn{2}{c}{68.70\%} \\
         Query success & \multicolumn{2}{c}{\textbf{94.96\%}} &  \multicolumn{2}{c}{76.85\%}\\ \hline
    \end{tabular}
}
\vspace{-1ex}
\caption{[Exp 1] Overall results (q-error and accuracy) for the test set comprising linear, 2-, and 3-way join queries.}
\label{tab:overall_error}
\vspace{-5ex}
\end{table}

\textbf{Predictions on the overall test set.} 
To evaluate the prediction quality of \costream for data beyond it was trained on, we first used our test data ($10\%$ of the full dataset), which has the same feature range as the training data shown in~\Cref{tab:training_data} but is unseen by the model.
It comprises of linear, 2-way, and 3-way join queries.
We report the overall prediction results for all our cost models in~\Cref{tab:overall_error}. 
We observe a median q-error of $1.33$ for throughput, $1.37$, and $1.46$, respectively, for end-to-end and processing latencies.
In addition, we achieved high accuracy for backpressure occurrence and query success of 87.89\% and 94.96\%.
In contrast, the baseline (flat vector) is much less precise and shows high q-errors (between 9.92 and 22.87) and lower accuracy (backpressure: 68.70\%; query success: 76.85\%).

\textbf{Predictions on heterogeneous hardware.} 
In the next step, we look into how well \costream{} predicts costs for heterogeneous hardware resources, as shown in \Cref{fig:qerror_over_hardware}. 
For this, we grouped the predictions over the mean of a specific hardware feature (like CPU or RAM) from all hosts that are part of a single query execution. 
For instance, we report the median q-error and prediction accuracy for all queries that are executed on hosts where CPU resources used for each operator lie in the same range (e.g., $[200\%, 300\%]$ refers to the case where an operator uses between two to three virtual CPU cores). 
Similarly, we group the prediction results over RAM, bandwidth, and latency of the computing nodes. 
As we can see, across hardware resources, we are achieving a median q-error of $1.6$ or better and an accuracy of above 85\%; i.e., the results are very accurate and stable across all different hardware dimensions. 

\textbf{Predictions results on the different query structures.} 
We investigated the prediction ability over different query structures on the test set. 
We show in \Cref{fig:qerror_over_query_type} how the q-error changes from simple and complex queries from left to right. 
For all regression tasks, we achieve a low q-error of below $1.6$, while the q-error for more complex queries is slightly higher as the cost estimation task becomes more difficult for them. 
More complex queries have a larger set of operators, plus their deployment on heterogeneous hardware makes cost estimation harder for \costream{}. 
Still, the model works precisely for all of the query types.
A similar behavior can be seen for query success and backpressure occurrence.

\vspace{-1ex}
\subsection{Exp 2: Placement optimization} \label{subsec:optimizer_performance}
\vspace{-1ex}

\begin{figure}
    \centering
    \vspace{-1ex}
    \includegraphics[height=2.5cm]{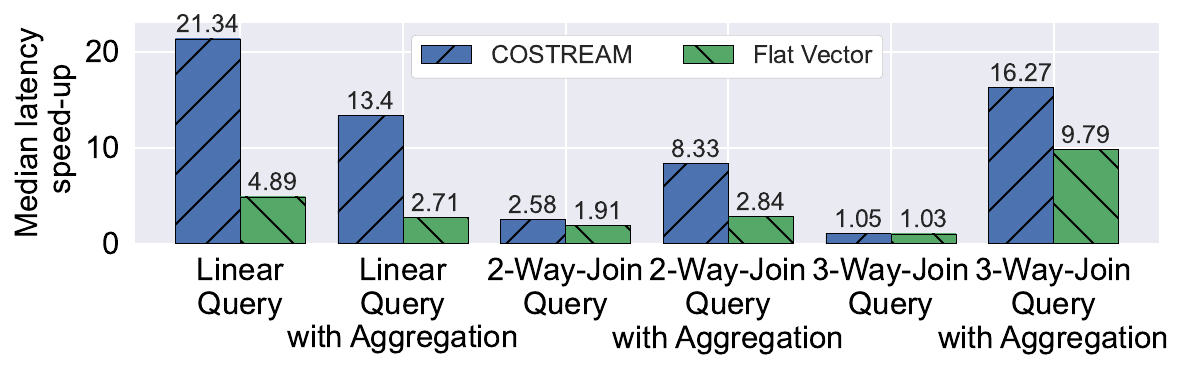}
    \vspace{-1ex}
    \caption{[Exp 2a] Median speed-ups for $L_p$ over different query types. 
    For each query type, the initial placements of 50 queries are optimized with estimates from \costream or the flat vector baseline and compared to an initial heuristic placement \cite{chaudhary2020}.
    We achieve high speed-ups with a median of up to 21.34$\times$.}
    \label{fig:optimizer_results}
    \vspace{-4ex}
\end{figure}

In the following, we evaluate the placement selection using \costream described in \Cref{sec:optimizer}. We present results where we select placement with an optimization objective as processing latency for the required initial placement.

[Exp 2a]: First, we applied the presented placement heuristics~\cite{chaudhary2020} to generate alternative placement candidates.
Then, we selected the best candidate using cost estimates given by \costream and the flat vector baseline.
The ratio between the latency of the initial and the best candidate placement is referred to as the \textit{speed-up factor}.
For \costream, we used three parallel latency models following the \textit{ensemble learning} approach to reduce the prediction uncertainty (cf.~\Cref{sec:optimizer}). 
For each query type (e.g., linear queries), we optimized $50$ queries with different complexities (e.g., filter predicates and event rates) and reported the median speed-up factors in \Cref{fig:optimizer_results}.
The results show that placement optimization using \costream improved the processing latency significantly for many queries.
Moreover, it clearly exceeds the speed-ups obtained by using the baseline.
For linear queries, a significant median speed-up factor of up to $21.34\times$ could be achieved with \costream, while the baseline achieves $4.89\times$.
Similarly, for the other more complex query types, high speed-ups could be reached, which shows that the initial placement optimization with \costream{} is highly beneficial.
In contrast, the flat vector baseline shows less accurate predictions and is therefore unable to find highly optimized placements.
Finding a good placement is highly important even for simple queries as these are long-running for days or even weeks.

\begin{figure}[t]
    \centering
    \includegraphics[width=1.0\linewidth]{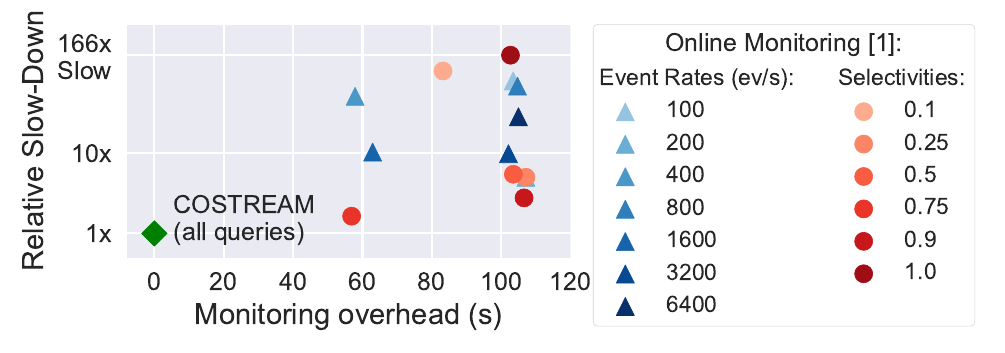}
    \vspace{-4ex}
    \caption{[Exp 2b] The relative slow-down factor (y-axis) in $L_p$ of a monitoring baseline~\cite{aniello2013} is up to $166\times$ in comparison to \costream that is the fastest (slow-down of 1) for all queries. Further, we report monitoring overhead of~\cite{aniello2013} (x-axis). This is the time needed for monitoring to find a placement that is competitive with the initial placement found by \costream{}, which can be up to $2$ minutes.}
    \vspace{-2ex}
    \label{fig:monitoring_queries}
\end{figure}

\begin{table}
    \centering
    \scalebox{0.71}{
    \begin{tabular}{lcccc}
    \hline
          \circles{\textsc{A}} & \textbf{\begin{tabular}[c]{@{}c@{}}RAM\\ (GB)\end{tabular}} & \textbf{\begin{tabular}[c]{@{}c@{}}CPU\\ (\% of a core)\end{tabular}} & \textbf{\begin{tabular}[c]{@{}c@{}}Bandwidth\\ (Mbit/s)\end{tabular}} & \textbf{\begin{tabular}[c]{@{}c@{}}Latency\\ (ms)\end{tabular}} \\ \hline
    Training Range & {\color[HTML]{009901} \begin{tabular}[c]{@{}c@{}}1, 2, 4, 8, \\ 16, 24, 32\end{tabular}} & {\color[HTML]{009901} \begin{tabular}[c]{@{}c@{}}50, 100, 200, \\ 300, 400, 500, \\ 600, 700, 800\end{tabular}} & {\color[HTML]{009901} \begin{tabular}[c]{@{}c@{}}25, 50, 100, 200, \\ 300, 800, 1600, \\ 3200, 4800, 800\end{tabular}} & {\color[HTML]{009901} \begin{tabular}[c]{@{}c@{}}1, 2, 5, 10, 20,\\  40, 80, 160\end{tabular}} \\
    Evaluation Range & {\color[HTML]{FFC702} \begin{tabular}[c]{@{}c@{}}1.5, 3, 6, \\ 12, 20, 28\end{tabular}} & {\color[HTML]{FFC702} \begin{tabular}[c]{@{}c@{}}75, 150, 250, \\ 350, 450, 550, \\ 650, 750\end{tabular}} & {\color[HTML]{FFC702} \begin{tabular}[c]{@{}c@{}}35, 75, 150, \\ 250, 550, 1200, \\ 1900, 4800, 8000\end{tabular}} & {\color[HTML]{FFC702} \begin{tabular}[c]{@{}c@{}}3,  7,  15, \\ 30, 60, 120\end{tabular}} \\ \hline
     & \multicolumn{1}{l}{} & \multicolumn{1}{l}{} & \multicolumn{1}{l}{} & \multicolumn{1}{l}{} \\
     \circles{\textsc{B}} & \multicolumn{2}{c}{\textbf{\costream}} & \multicolumn{2}{c}{\textbf{Flat Vector}} \\ \hline
    \textbf{Metric} & \textbf{Q50} & Q95 & \textbf{Q50} & Q95 \\ \hline
    Throughput & \textbf{1.37} & 8.28 & \textbf{15.63} & 282.50 \\
    E2E-Latency & \textbf{1.59} & 25.33 & \textbf{63.79} & 869.85 \\
    Processing Latency & \textbf{1.54} & 17.78 & \textbf{27.85} & 282.50 \\ \hline
    Backpressure & \multicolumn{2}{c}{\textbf{88.04\%}} & \multicolumn{2}{c}{72.83\%} \\
    Query Success & \multicolumn{2}{c}{\textbf{87.13\%}} & \multicolumn{2}{c}{68.32\%} \\ \hline
    \end{tabular}
    }
    \vspace{-1ex}
    \caption{[Exp 3] \circles{\textsc{A}} Evaluation range for $n=100$ queries that are within the training data range but differ from it. \circles{\textsc{B}} Interpolation results (q-error and accuracy) for queries running on entirely unseen hardware resources. 
    \costream{} can predict precisely even for unseen hardware configurations.}
    \label{tab:interpolation_results}
    \vspace{-6ex}
\end{table}

\begin{table*}
\begin{minipage}{\columnwidth}
\scalebox{0.7}{
\centering
\begin{tabular}{lcccccccc}
\multicolumn{9}{c}{\textbf{\circles{\textsc{A}} Extrapolation towards stronger resources}} \\
\hline
 & \multicolumn{2}{c}{\textbf{\begin{tabular}[c]{@{}c@{}}RAM\\ (GB)\end{tabular}}} & \multicolumn{2}{c}{\textbf{\begin{tabular}[c]{@{}c@{}}CPU\\ (\% of a core)\end{tabular}}} & \multicolumn{2}{c}{\textbf{\begin{tabular}[c]{@{}c@{}}Bandwidth\\ (Mbit/s)\end{tabular}}} & \multicolumn{2}{c}{\textbf{\begin{tabular}[c]{@{}c@{}}Latency\\ (ms)\end{tabular}}} \\ \hline
Training Range & \multicolumn{2}{c}{{\color[HTML]{009901} 1, 2, 4, 8, 16}} & \multicolumn{2}{c}{{\color[HTML]{009901} \begin{tabular}[c]{@{}c@{}}50, 100, 200, 300,\\ 400, 500, 600\end{tabular}}} & \multicolumn{2}{c}{{\color[HTML]{009901} \begin{tabular}[c]{@{}c@{}}25, 50, 100, 200, \\ 300, 800,  1.6k, 3.2k\end{tabular}}} & \multicolumn{2}{c}{{\color[HTML]{009901} \begin{tabular}[c]{@{}c@{}}5, 10, 20, 40,\\ 80, 160\end{tabular}}} \\
Evaluation Range & \multicolumn{2}{c}{{\color[HTML]{F8A102} 24, 32}} & \multicolumn{2}{c}{{\color[HTML]{F8A102} 700, 800}} & \multicolumn{2}{c}{{\color[HTML]{F8A102} 64k, 10k}} & \multicolumn{2}{c}{{\color[HTML]{F8A102} 1, 2}} \\ \hline
\textbf{Metric} & \textbf{Q50} & Q95 & \textbf{Q50} & Q95 & \textbf{Q50} & Q95 & \textbf{Q50} & Q95 \\ \hline
Throughput & \textbf{1.66} & 5.88 & \textbf{1.72} & 9.40 & \textbf{1.48} & 6.55 & \textbf{1.52} & 5.60 \\
E2E-Latency & \textbf{1.85} & 29.08 & \textbf{1.67} & 9.43 & \textbf{1.75} & 17.18 & \textbf{3.55} & 30.90 \\
Processing Latency & \textbf{1.88} & 11.32 & \textbf{1.75} & 6.81 & \textbf{1.63} & 13.89 & \textbf{3.83} & 19.43 \\ \hline
Backpressure & \multicolumn{2}{c}{\textbf{85.37\%}} & \multicolumn{2}{c}{\textbf{86.59\%}} & \multicolumn{2}{c}{\textbf{86.59\%}} & \multicolumn{2}{c}{\textbf{88.89\%}} \\
Query Success & \multicolumn{2}{c}{\textbf{77.00\%}} & \multicolumn{2}{c}{\textbf{93.14\%}} & \multicolumn{2}{c}{\textbf{87.25\%}} & \multicolumn{2}{c}{\textbf{92.93\%}} \\ \hline
\end{tabular}
}
\end{minipage}\hfill
\begin{minipage}{\columnwidth}
\scalebox{0.7}{
\begin{tabular}{lcccccccc}
\multicolumn{9}{c}{\textbf{\circles{\textsc{B}} Extrapolation towards weaker resources}} \\
\hline
 & \multicolumn{2}{c}{\textbf{\begin{tabular}[c]{@{}c@{}}RAM\\ (GB)\end{tabular}}} & \multicolumn{2}{c}{\textbf{\begin{tabular}[c]{@{}c@{}}CPU\\ (\% of a core)\end{tabular}}} & \multicolumn{2}{c}{\textbf{\begin{tabular}[c]{@{}c@{}}Bandwidth\\ (Mbit/s)\end{tabular}}} & \multicolumn{2}{c}{\textbf{\begin{tabular}[c]{@{}c@{}}Latency\\ (ms)\end{tabular}}} \\ \hline
Training Range & \multicolumn{2}{c}{{\color[HTML]{009901} \begin{tabular}[c]{@{}c@{}}4, 8, 16, \\ 24, 32\end{tabular}}} & \multicolumn{2}{c}{{\color[HTML]{009901} \begin{tabular}[c]{@{}c@{}}200, 300, 400, \\ 500, 600, 700, 800\end{tabular}}} & \multicolumn{2}{c}{{\color[HTML]{009901} \begin{tabular}[c]{@{}c@{}}100, 200, 300, 800, \\ 1.6k, 3.2k, 6.4k, 10k\end{tabular}}} & \multicolumn{2}{c}{{\color[HTML]{009901} \begin{tabular}[c]{@{}c@{}}1,2,5,10\\ 20, 40\end{tabular}}} \\
Evaluation Range & \multicolumn{2}{c}{{\color[HTML]{F8A102} 1, 2}} & \multicolumn{2}{c}{{\color[HTML]{F8A102} 50, 100}} & \multicolumn{2}{c}{{\color[HTML]{F8A102} 25, 50}} & \multicolumn{2}{c}{{\color[HTML]{F8A102} 80, 160}} \\ \hline
\textbf{Metric} & \textbf{Q50} & Q95 & \textbf{Q50} & Q95 & \textbf{Q50} & Q95 & \textbf{Q50} & Q95 \\ \hline
Throughput & \textbf{1.79} & 7.60 & \textbf{1.61} & 13.16 & \textbf{1.42} & 5.30 & \textbf{3.25} & 33.65 \\
E2E-Latency & \textbf{1.72} & 13.69 & \textbf{2.75} & 111.53 & \textbf{1.46} & 5.30 & \textbf{2.10} & 54.13 \\
Processing Latency & \textbf{1.49} & 13.27 & \textbf{2.96} & 77.56 & \textbf{1.68} & 12.94 & \textbf{6.09} & 406.83 \\ \hline
Backpressure & \multicolumn{2}{c}{\textbf{91.03\%}} & \multicolumn{2}{c}{\textbf{75.00\%}} & \multicolumn{2}{c}{\textbf{91.92\%}} & \multicolumn{2}{c}{\textbf{67.82\%}} \\
Query Success & \multicolumn{2}{c}{\textbf{78.79\%}} & \multicolumn{2}{c}{\textbf{86.67\%}} & \multicolumn{2}{c}{\textbf{92.59\%}} & \multicolumn{2}{c}{\textbf{74.51\%}} \\ \hline
\end{tabular}
}
\end{minipage}
\vspace{-1ex}
\caption{[Exp 4] Extrapolation results (q-error and accuracy) towards stronger \circles{\textsc{A}} and weaker \circles{\textsc{B}} hardware and network resources. For each dimension, \costream{} was trained on a reduced training range and evaluated with $n=100$ queries out of the unseen evaluation range. \costream{} can predict precisely for hardware properties beyond the initial training range for both stronger and weaker resources.}
\vspace{-2ex}
\label{tab:extrapolation_results}
\end{table*}

\begin{table*}
\scalebox{0.54}{
\begin{tabular}{lcccccccccccclcccccccccccccccl}
 & \multicolumn{12}{c}{\textbf{\circles{\textsc{A}} [Exp 5] Unseen query pattern}} &  & \multicolumn{16}{c}{\textbf{\circles{\textsc{B}} [Exp 6] Unseen benchmarks}} \\ \cline{2-13} \cline{15-30} 
 & \multicolumn{4}{c|}{\textbf{2-Fiter Chain}} & \multicolumn{4}{c|}{\textbf{3-Filter-Chain}} & \multicolumn{4}{c}{\textbf{4-Filter-Chain}} &  & \multicolumn{4}{c|}{\textbf{Advertisement}} & \multicolumn{4}{c|}{\textbf{Spike Detection}} & \multicolumn{4}{c|}{\textbf{\begin{tabular}[c]{@{}c@{}}Smart Grid\\ (global)\end{tabular}}} & \multicolumn{4}{c}{\textbf{\begin{tabular}[c]{@{}c@{}}Smart Grid \\ (local)\end{tabular}}} \\ \cline{2-13} \cline{15-30} 
 & \multicolumn{2}{c}{\costream} & \multicolumn{2}{c|}{\flatvector} & \multicolumn{2}{c}{\costream} & \multicolumn{2}{c|}{\flatvector} & \multicolumn{2}{c}{\costream} & \multicolumn{2}{c}{\flatvector} & \textbf{} & \multicolumn{2}{c}{\costream} & \multicolumn{2}{c|}{\flatvector} & \multicolumn{2}{c}{\costream} & \multicolumn{2}{c|}{\flatvector} & \multicolumn{2}{c}{\costream} & \multicolumn{2}{c|}{\flatvector} & \multicolumn{2}{c}{\costream} & \multicolumn{2}{c}{\flatvector} \\ \cline{1-13} \cline{15-30} 
Metric & Q50 & Q95 & Q50 & \multicolumn{1}{c|}{Q95} & Q50 & Q95 & Q50 & \multicolumn{1}{c|}{Q95} & Q50 & Q95 & Q50 & Q95 & \textbf{} & Q50 & Q95 & Q50 & \multicolumn{1}{c|}{Q95} & Q50 & Q05 & Q50 & \multicolumn{1}{c|}{Q95} & Q50 & Q95 & Q50 & \multicolumn{1}{c|}{Q95} & Q50 & Q95 & Q50 & \multicolumn{1}{c}{Q95} \\ \cline{1-13} \cline{15-30} 
Throughput & \textbf{2.74} & 64.35 & 5.52 & \multicolumn{1}{c|}{244.38} & \textbf{2.87} & 75.29 & 18.82 & \multicolumn{1}{c|}{1078.26} & \textbf{5.51} & 445.87 & 82.71 & 3672.13 &  & \textbf{1.98} & 11.01 & 3.12 & \multicolumn{1}{c|}{46.11} & \textbf{3.67} & 66.48 & 274.04 & \multicolumn{1}{c|}{891.99} & \textbf{1.44} & 5.98 & 104.79 & \multicolumn{1}{c|}{106.06} & \textbf{1.43} & 10.51 & 104.79 & 106.12 \\
E2E-Latency & \textbf{1.68} & 21.81 & 259.98 & \multicolumn{1}{c|}{2302.38} & \textbf{2.15} & 11.81 & 536.38 & \multicolumn{1}{c|}{1855.05} & \textbf{2.68} & 23.99 & 538.10 & 1877.68 &  & 2.02 & 15.08 & \textbf{1.32} & \multicolumn{1}{c|}{40.59} & \textbf{1.41} & 17.55 & 2.28 & \multicolumn{1}{c|}{1017.96} & \textbf{2.01} & 50.17 & 118.77 & \multicolumn{1}{c|}{639.79} & \textbf{1.67} & 31.00 & 143.22 & 669.20 \\
Proc-Latency & \textbf{1.69} & 48.26 & 48.93 & \multicolumn{1}{c|}{341.70} & \textbf{1.64} & 5.41 & 63.62 & \multicolumn{1}{c|}{266.80} & \textbf{1.61} & 5.38 & 55.27 & 270.36 &  & \textbf{2.27} & 15.01 & 3.62 & \multicolumn{1}{c|}{41.37} & \textbf{1.63} & 12.92 & 5.32 & \multicolumn{1}{c|}{339.82} & \textbf{1.48} & 12.70 & 35.48 & \multicolumn{1}{c|}{161.60} & \textbf{1.54} & 7.96 & 37.57 & 174.38 \\ \cline{1-13} \cline{15-30} 
Backpressure & \multicolumn{2}{c}{\textbf{88\%}} & \multicolumn{2}{c|}{68\%} & \multicolumn{2}{c}{\textbf{85\%}} & \multicolumn{2}{c|}{79\%} & \multicolumn{2}{c}{\textbf{82\%}} & \multicolumn{2}{c}{79\%} &  & \multicolumn{2}{c}{\textbf{85\%}} & \multicolumn{2}{c|}{80\%} & \multicolumn{2}{c}{\textbf{78\%}} & \multicolumn{2}{c|}{55\%} & \multicolumn{2}{c}{\textbf{81\%}} & \multicolumn{2}{c|}{29\%} & \multicolumn{2}{c}{\textbf{86\%}} & \multicolumn{2}{c}{23\%} \\
Query success & \multicolumn{2}{c}{\textbf{100\%}} & \multicolumn{2}{c|}{4\%} & \multicolumn{2}{c}{\textbf{100\%}} & \multicolumn{2}{c|}{6\%} & \multicolumn{2}{c}{\textbf{100\%}} & \multicolumn{2}{c}{6\%} &  & \multicolumn{2}{c}{100\%} & \multicolumn{2}{c|}{100\%} & \multicolumn{2}{c}{100\%} & \multicolumn{2}{c|}{0\%} & \multicolumn{2}{c}{\textbf{100\%}} & \multicolumn{2}{c|}{100\%} & \multicolumn{2}{c}{\textbf{100\%}} & \multicolumn{2}{c}{100\%} \\ \cline{1-13} \cline{15-30} 
\end{tabular}
}
\vspace{-1ex}
 \caption{\circles{\textsc{A}} [Exp 5a] Prediction results (q-error and accuracy) for queries that are unseen in the training data in terms of their structure.
 The results are in an acceptable range but decrease with increasing complexity. Model fine-tuning can be applied to improve the results.\\\hspace{\textwidth}
  \circles{\textsc{B}} [Exp 6] Results (q-error and accuracy) for benchmark queries from \cite{bordin_dspbench_2020}. We executed each query $n=100$ times with different event rates and operator placements.  For these benchmarks with unseen data distribution, \costream{} predicts cost precisely.}
  \label{tab:unseen_structures}
\vspace{-5ex}
\end{table*}

[Exp 2b]:
We additionally compare \costream against an online monitoring approach that can be integrated into Storm (based on \cite{aniello2013, eskandari2019}). 
We show the performance of a linear filter query over varied selectivities as well as input event rates. The results are shown in \Cref{fig:monitoring_queries}.
This approach initially uses a heuristic for placement, which is comparable to our heuristic baseline in the previous experiment. 
In monitoring, after the query execution stabilizes, a re-deployment is triggered based on collected runtime statistics (e.g., CPU utilization and network usage). We report two metrics to show the advantages of \costream over the baseline as seen in \Cref{fig:monitoring_queries}:

(1) \textit{Relative slow-down}: While \costream directly starts with a placement that aims to minimize the processing latency, the baseline starts with placement based on a heuristic that comes with higher latencies.
We report the initial relative latency difference ($L_p$) as a slow-down factor measured as the ratio of processing latency achieved using baseline over that of \costream. 
As seen in the y-axis in \Cref{fig:monitoring_queries}, the baseline approach is up to $166\times$ slower. Moreover, the initial placement found by \costream{} is better across all queries.

(2) \textit{Monitoring overhead}: Online monitoring approaches need to monitor the execution using runtime statistics and then migrate operators during execution to adjust the deployment.
However, adjusting the deployment has high overheads since operators and their execution state (e.g., windows) need to be migrated between machines. 
As a second interesting metric, we thus report how much time the monitoring approach needs to re-adjust the initial placement and find a more optimal placement that is competitive with the initial placement found by \costream{} (i.e., the processing latency is the same or slightly better).
The time to find such competitive deployments (called monitoring overhead) is shown on the x-axis in \Cref{fig:monitoring_queries}.
We see that the monitoring overhead ranges between 70 seconds and is often even more than \textit{two} minutes for several queries.
This overhead does not occur when using \costream.

\vspace{-1.5ex}
\subsection{Exp 3: Generalization over hardware (interpolation)}
\vspace{-0.5ex}
In this experiment, we show how \costream{} generalized for hardware characteristics that are unseen during the training.
While our model is trained with the hardware features from \Cref{tab:training_data}, we generated and evaluated a new, unseen test set out of 100 test queries that were executed on hardware that differed from the training set but lies \emph{within} the training range.
\Cref{tab:interpolation_results} contains these ranges \circles{A} and the overall results \circles{B} for this unseen interpolation test set.
\costream achieves high accuracy for the generalization experiment, which is important for placement decisions on unseen hardware.
Median q-errors are between $1.37$ and $1.59$, and accuracy achieves up to $88.04\%$.
Moreover, \costream{} outperforms the flat vector baseline for all cost metrics which justifies our model architecture, which was explicitly developed to accurately enable generalizable cost predictions on heterogeneous hardware.

\vspace{-0.5ex}
\subsection{Exp 4: Generalization over hardware (extrapolation)}
\vspace{-0.5ex}
Even more relevant and challenging is to predict costs for hardware resources that are \textit{beyond} the initial training range.
In this experiment, we evaluated how \costream predicts costs for either weaker or stronger resources beyond the initial range.
For instance, query executions with larger RAM configurations from the training dataset\footnote{In this experiment we trained \costream with a restricted training data range to test extrapolation.} were used to train a model, and then predictions were generated for smaller amounts of RAM.
Similarly, this was repeated for CPU, network bandwidth, and latency. 
The results are presented in \Cref{tab:extrapolation_results} \circles{A} for stronger and \circles{B} weaker resources, showing that for this more challenging scenario \costream{} can predict costs for \textit{unseen} hardware and network resources beyond the initial training data range.
Particularly for CPU and RAM, we see that our model still is highly accurate.
We want to note that the extrapolation results for unseen network latencies are not as good, with a median q-error of up to $6.09$ for higher network latencies (i.e., slower networks). 
However, it is important that the latencies for testing are $4\times$ as high as the training range. 

\subsection{Exp 5: Unseen query patterns} \label{subsec:unseen_patterns}
[Exp 5a] We further investigated how \costream{} predicts for queries that use query patterns \textit{unseen} in the training set. 
Modern \ac{dsps} are typically required to define and wire the query operators by themselves, as streaming query languages have not yet been widely adopted. 
This opens up an infinite space for query patterns beyond structures included in our dataset. We investigate how \costream{} predicts these as it has to face even unseen query patterns during operation.
Precisely, we created and executed longer filter-chain queries unseen during training. Unseen filter chains use $2$, $3$, or $4$ filter operators with random filter properties, while training has only seen 1 subsequent filter operator.
We report the results in \Cref{tab:unseen_structures} \circles{A}. 
In general, it can be seen that the model accuracy is still accurate with median q-errors of up to $1.68$ for 2-filter chains.
For more filters, the prediction quality slightly decreases, especially for the tail q-errors.
Moreover, important is that \costream{} outperforms the flat vector model, where we generally observed much higher q-errors.
For query success prediction, the baseline (flat vector) is in particular of low quality. 
We analyzed this and found that all queries are classified by the baseline as failing if they include more than one filter. 
This shows that the baseline unable to extrapolate to unseen query patterns, proving our model design.

\begin{figure}
    \centering
    \vspace{-1ex}
    \includegraphics[width=\columnwidth]{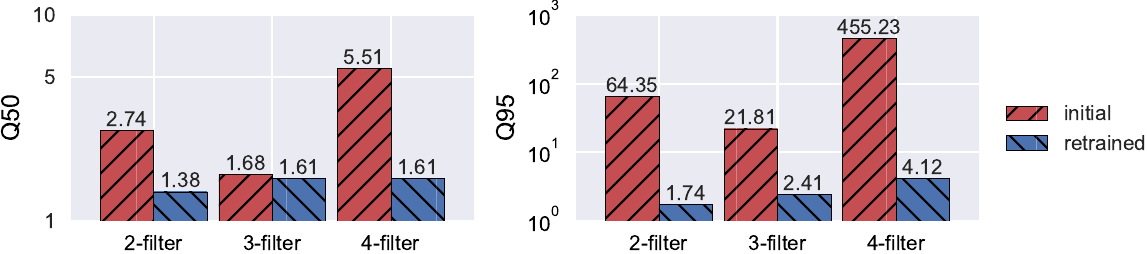}
    \vspace{-4ex}
    \caption{[Exp 5b] Prediction results for $T$ \textit{before} and \textit{after} applying fine-tuning for unseen query structures, which improves the results while requiring only a small amount of additional data.}
    \label{fig:retraining}
    \vspace{-4.5ex}
\end{figure}

[Exp 5b] A way to improve the \costream{} results for unseen query patterns is to apply \textit{few-shot learning}.
This means to train the model on a small number of additional queries of interest. 
To demonstrate this, we tune our throughput model with only $3000$ additional filter queries and present the improved results in \Cref{fig:retraining}. 
Fine-tuning was particularly beneficial for $3$- and $4$-filter chains, where the q-errors decreased significantly (e.g. $5.51$ to $1.61$ for 4-filter-chain).
Fine-tuning can also be applied to support entirely unseen operators that haven't been part of the training set.

\begin{figure*}
\begin{minipage}[t]{0.48\linewidth}
    \vspace{-\topskip}
    \centering
    \includegraphics[height=2.2cm]{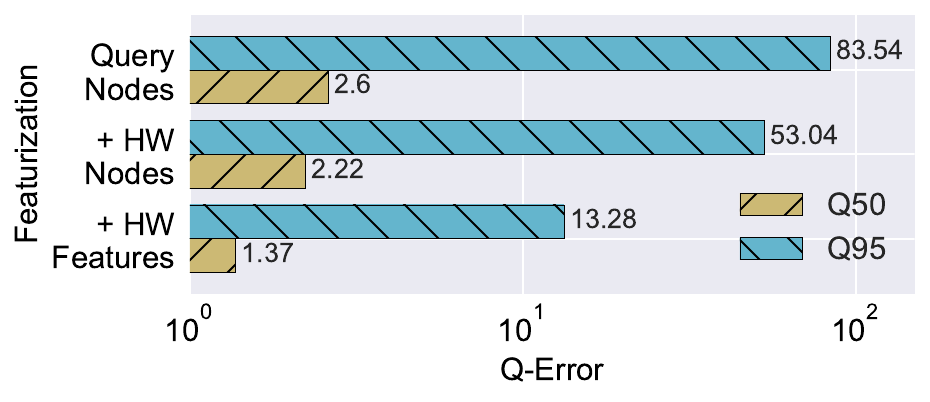}
    \vspace{-1ex}
    \caption{[Exp 7a] Prediction results (q-error) for $L_e$ of different featurization schemes. 
    The upper scheme does not consider placement or hardware at all, while the middle scheme includes hardware nodes and thus models the operator placement. 
    The full featurization (bottom) shows the best results.}
    \label{fig:ablation_study}
    \vspace{-4ex}
\end{minipage}\hfill
\begin{minipage}[t]{0.48\linewidth}
    \vspace{-\topskip}
    \centering
    \includegraphics[scale=0.52]{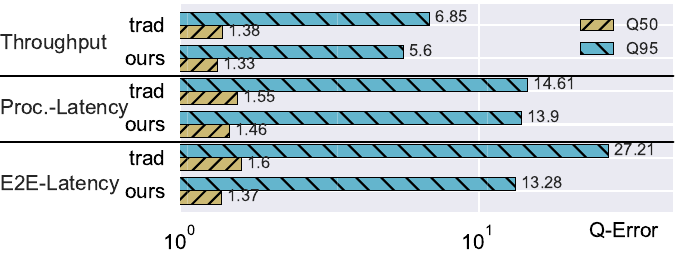}
    \vspace{-1ex}
    \caption{[Exp 7b] Prediction results (q-error) of a traditional message passing scheme vs ours. Our novel message-passing scheme shows better results and is thus beneficial for precise cost estimation.}
    \label{fig:message_passing_ablation}
    \vspace*{-5ex}
\end{minipage}
\end{figure*}

\subsection{Exp 6: Unseen benchmarks}
\vspace{-1ex}
In this experiment, we apply \costream{} on real-world queries from \cite{bordin_dspbench_2020}, that the model has not seen during training. 
The main challenge with the selected queries for our cost model lies in the different data distribution, which is at the heart of the data streams.
While our benchmark workloads are generated, these real-world benchmarks come with different, realistic data distributions.
However, published benchmarks heavily rely on user-defined operators \cite{hesse_espbench_2021, shukla2017, lu2014}, that are not yet applicable for \costream. We excluded such queries.

\textit{Advertisement benchmark}: In this benchmark, the ratio of aggregates of a click stream and an impression stream is calculated, which are joined and grouped by two attributes. 
The initial query \cite{bordin_dspbench_2020} is complex and cannot easily be expressed into algebraic operators. 
Thus, we use a sub-query with two streams, a filter, and a windowed join. 
The data is real-world.

\textit{Spike detection benchmark}: This real-world benchmark query is motivated by an \ac{iot}-use case. 
The target is to filter out spikes of an incoming sensor data stream. 

\textit{Smart grid benchmark}: This benchmark was published in the DEBS Grand Challenge 2014 \cite{jerzak2014} and came along with various sub-queries \cite{koliousis2016}. 
In our work, we consider parts of the outlier detection task and implement a sliding window query on the incoming stream (global) to compute the global energy consumption. 
Furthermore, we implemented a query that computes the local energy consumption by grouping previous results over the corresponding households. 
The data is generated using the implementation from \cite{bordin_dspbench_2020}.
As no source event rates were given in these benchmarks, we executed each benchmark query 100 times with random event rates and different placements.
By the results from \Cref{tab:unseen_structures} \circles{B}, it becomes clear that \costream{} can again precisely predict the query costs of unseen benchmarks highly accurately with a median q-error between $1.41$ and $3.67$. 
In contrast, the baseline (flat vector) again shows much higher prediction errors. 
The main reason is again that the baseline does not generalize well to unseen workloads as previously discussed (cf. \Cref{subsec:unseen_patterns}).
For example, as the spike detection benchmark contains queries with two filters, the baseline again fails to predict query success and throughput.
Moreover, the Smart Grid queries contain an unseen window length for that \costream{} can extrapolate successfully.

\vspace{-1ex}
\subsection{Exp 7: Ablation studies}
\vspace{-1ex}
In the following, we evaluate different design variations. 

[Exp 7a] \textit{Feature ablation.}
At first, we investigate our selection of features.
(1) A naive approach would be to encode only the query operators and data sources/sinks but omit hardware nodes entirely. 
This way, the model would only know the query logic and not the operator placement and hardware configurations.
(2) A more sophisticated approach additionally includes the placement and the co-location of operators but does not know about the hardware and network resources.
(3) We compare both alternatives to our full featurization scheme for predicting $L_e$ and show the results in \Cref{fig:ablation_study}.
Our full scheme has the most accurate predictions with a median q-error of $1.37$ while using only operator nodes leads to a lower q-error of $2.6$.
Adding at least the hardware nodes returns a median q-error of $2.22$. 
Thus, the operator placement and hardware features add important information.

[Exp 7b] \textit{Message passing ablation.}
Moreover, we compare our novel message-passing scheme with a traditional scheme, where in each epoch all graph nodes are updated with the messages from their neighbors, regardless of their node type. 
In \Cref{fig:message_passing_ablation}, we demonstrate that our scheme compared to a traditional scheme yields higher prediction accuracy across all regression tasks, highlighting its benefit for cost estimation.
\section{Related work}
\label{sec:related_work}

\textbf{Analytical and heuristic approaches.} 
Close to our work is R-Storm, which tries to match the resource needs of streaming operators to given resources via monitoring~\cite{peng2015}. 
Similarly, other works rely on monitoring for operator placement~\cite{xu2014}, including recent work that targets the co-location of streaming operators explicitly~\cite{liu2023}. 
Some works combine monitoring with heuristics, such as~\cite{aniello2013}, that, however, does not take hardware heterogeneity into account.
Other approaches rely on meta- or custom-heuristics~\cite{eskandari2019, chandramouli2011}. 
\cite{nardelli2019} proposes a set of 
heuristics to solve the operator placement
but does not model the query logic or heterogeneous hardware. 
Apache Flink~\cite{carbone2015} uses a heuristic-based optimization algorithm that is built upon \cite{alexandrov2014}.
However, these approaches require monitoring or runtime statistics and thus initial query optimization is not possible.

\textbf{Learned approaches.}
Our previous work \cite{heinrich2022} proposed a learned cost model for \ac{dsps} queries but did not take heterogeneous operator placement into account.
\cite{foroni2018} is close to this work, predicting query execution costs with regression models but relies on monitoring input.
To predict application latency, \cite{li2016} introduces various features that model hardware resources.
\cite{imai2017} proposes two throughput models that assume knowledge about internal processing times, while~\cite{wang2017} takes the query and hardware properties into account, which, however, are not heterogeneous 
and not transferable.
Other approaches optimize for operator placement by applying methods of machine learning~\cite{luthra2021, luthra2018, li2018, mao2019}.
However, they either lack generalizability to unseen query workloads, make use of monitoring information, or assume hardware homogeneity.
\section{Conclusion and outlook}
\vspace{-1ex}
\label{sec:conclusion}

In this paper, we presented \costream, a cost model that predicts throughput, end-to-end latency, processing latency, query success, and the backpressure occurrence of a \ac{dsps} query to be executed on heterogeneous hardware.
We further demonstrated how to use \costream{} as an important component for solving the initial operator placement problem in \ac{iot}-scenarios. 
There are various promising ways to extend our work. 
A natural extension would be to extensively apply learned cost models for \ac{dsps} on various other optimization problems, like the elasticity or the parallelism tuning problem~\cite{agnihotri2023} or even a generic cost model for several streaming optimizations. Our proposed graph structure is adaptable to all of these extensions.
Other interesting research directions are making \costream generalizable across different \ac{dsps} like Flink and Spark and extending \costream for metrics related to cloud deployments like predicting monetary costs.

\section*{Acknowledgements}
\footnotesize{This work has been supported by the IPF program at DHBW Mannheim, the Collaborative Research Center (CRC) 1053-MAKI funded by DFG, the NHR4CES program, hessian.AI at TU Darmstadt as well as DFKI Darmstadt.}

\bibliographystyle{IEEEtran}
\balance
\bibliography{bibliography.bib}

\begin{thebibliography}{10}
\providecommand{\url}[1]{#1}
\csname url@samestyle\endcsname
\providecommand{\newblock}{\relax}
\providecommand{\bibinfo}[2]{#2}
\providecommand{\BIBentrySTDinterwordspacing}{\spaceskip=0pt\relax}
\providecommand{\BIBentryALTinterwordstretchfactor}{4}
\providecommand{\BIBentryALTinterwordspacing}{\spaceskip=\fontdimen2\font plus
\BIBentryALTinterwordstretchfactor\fontdimen3\font minus \fontdimen4\font\relax}
\providecommand{\BIBforeignlanguage}[2]{{%
\expandafter\ifx\csname l@#1\endcsname\relax
\typeout{** WARNING: IEEEtran.bst: No hyphenation pattern has been}%
\typeout{** loaded for the language `#1'. Using the pattern for}%
\typeout{** the default language instead.}%
\else
\language=\csname l@#1\endcsname
\fi
#2}}
\providecommand{\BIBdecl}{\relax}
\BIBdecl

\bibitem{aniello2013}
\BIBentryALTinterwordspacing
L.~Aniello, R.~Baldoni, and L.~Querzoni, ``Adaptive online scheduling in storm,'' in \emph{The 7th {ACM} International Conference on Distributed Event-Based Systems, {DEBS} '13, Arlington, TX, {USA} - June 29 - July 03, 2013}, S.~Chakravarthy, S.~D. Urban, P.~R. Pietzuch, and E.~A. Rundensteiner, Eds.\hskip 1em plus 0.5em minus 0.4em\relax ACM, 2013, pp. 207--218. [Online]. Available: \url{https://doi.org/10.1145/2488222.2488267}
\BIBentrySTDinterwordspacing

\bibitem{imai2017}
\BIBentryALTinterwordspacing
S.~Imai, S.~Patterson, and C.~A. Varela, ``Maximum sustainable throughput prediction for data stream processing over public clouds,'' in \emph{Proceedings of the 17th {IEEE/ACM} International Symposium on Cluster, Cloud and Grid Computing, {CCGRID} 2017, Madrid, Spain, May 14-17, 2017}.\hskip 1em plus 0.5em minus 0.4em\relax {IEEE} Computer Society / {ACM}, 2017, pp. 504--513. [Online]. Available: \url{https://doi.org/10.1109/CCGRID.2017.105}
\BIBentrySTDinterwordspacing

\bibitem{chandramouli2011}
\BIBentryALTinterwordspacing
B.~Chandramouli, J.~Goldstein, R.~S. Barga, M.~Riedewald, and I.~Santos, ``Accurate latency estimation in a distributed event processing system,'' in \emph{Proceedings of the 27th International Conference on Data Engineering, {ICDE} 2011, April 11-16, 2011, Hannover, Germany}, S.~Abiteboul, K.~B{\"{o}}hm, C.~Koch, and K.~Tan, Eds.\hskip 1em plus 0.5em minus 0.4em\relax {IEEE} Computer Society, 2011, pp. 255--266. [Online]. Available: \url{https://doi.org/10.1109/ICDE.2011.5767926}
\BIBentrySTDinterwordspacing

\bibitem{heinrich2022}
\BIBentryALTinterwordspacing
R.~Heinrich, M.~Luthra, H.~Kornmayer, and C.~Binnig, ``Zero-shot cost models for distributed stream processing,'' in \emph{16th {ACM} International Conference on Distributed and Event-based Systems, {DEBS} 2022, Copenhagen, Denmark, June 27 - 30, 2022}, Y.~Zhou, P.~K. Chrysanthis, V.~Gulisano, and E.~T. Zacharatou, Eds.\hskip 1em plus 0.5em minus 0.4em\relax ACM, 2022, pp. 85--90. [Online]. Available: \url{https://doi.org/10.1145/3524860.3539639}
\BIBentrySTDinterwordspacing

\bibitem{pietzuch2006}
\BIBentryALTinterwordspacing
P.~R. Pietzuch, J.~Ledlie, J.~Shneidman, M.~Roussopoulos, M.~Welsh, and M.~I. Seltzer, ``Network-aware operator placement for stream-processing systems,'' in \emph{Proceedings of the 22nd International Conference on Data Engineering, {ICDE} 2006, 3-8 April 2006, Atlanta, GA, {USA}}, L.~Liu, A.~Reuter, K.~Whang, and J.~Zhang, Eds.\hskip 1em plus 0.5em minus 0.4em\relax {IEEE} Computer Society, 2006, p.~49. [Online]. Available: \url{https://doi.org/10.1109/ICDE.2006.105}
\BIBentrySTDinterwordspacing

\bibitem{wang2017}
\BIBentryALTinterwordspacing
C.~Wang, X.~Meng, Q.~Guo, Z.~Weng, and C.~Yang, ``Automating characterization deployment in distributed data stream management systems,'' \emph{{IEEE} Trans. Knowl. Data Eng.}, vol.~29, no.~12, pp. 2669--2681, 2017. [Online]. Available: \url{https://doi.org/10.1109/TKDE.2017.2751606}
\BIBentrySTDinterwordspacing

\bibitem{cardellini2016}
\BIBentryALTinterwordspacing
V.~Cardellini, V.~Grassi, F.~L. Presti, and M.~Nardelli, ``Optimal operator placement for distributed stream processing applications,'' in \emph{Proceedings of the 10th {ACM} International Conference on Distributed and Event-based Systems, {DEBS} '16, Irvine, CA, USA, June 20 - 24, 2016}, A.~Gal, M.~Weidlich, V.~Kalogeraki, and N.~Venkasubramanian, Eds.\hskip 1em plus 0.5em minus 0.4em\relax ACM, 2016, pp. 69--80. [Online]. Available: \url{https://doi.org/10.1145/2933267.2933312}
\BIBentrySTDinterwordspacing

\bibitem{nardelli2019}
\BIBentryALTinterwordspacing
M.~Nardelli, V.~Cardellini, V.~Grassi, and F.~L. Presti, ``Efficient operator placement for distributed data stream processing applications,'' \emph{{IEEE} Trans. Parallel Distributed Syst.}, vol.~30, no.~8, pp. 1753--1767, 2019. [Online]. Available: \url{https://doi.org/10.1109/TPDS.2019.2896115}
\BIBentrySTDinterwordspacing

\bibitem{luthra2021}
\BIBentryALTinterwordspacing
M.~Luthra, B.~Koldehofe, N.~Danger, P.~Weisenburger, G.~Salvaneschi, and I.~Stavrakakis, ``{TCEP:} transitions in operator placement to adapt to dynamic network environments,'' \emph{J. Comput. Syst. Sci.}, vol. 122, pp. 94--125, 2021. [Online]. Available: \url{https://doi.org/10.1016/j.jcss.2021.05.003}
\BIBentrySTDinterwordspacing

\bibitem{liu2023}
\BIBentryALTinterwordspacing
F.~Liu, W.~Zhu, W.~Mu, Y.~Zhang, M.~Li, C.~Ma, and W.~Wang, ``Online runtime environment prediction for complex colocation interference in distributed streaming processing,'' in \emph{Computational Science - {ICCS} 2023 - 23rd International Conference, Prague, Czech Republic, July 3-5, 2023, Proceedings, Part {II}}, ser. Lecture Notes in Computer Science, J.~Mikyska, C.~de~Mulatier, M.~Paszynski, V.~V. Krzhizhanovskaya, J.~J. Dongarra, and P.~M.~A. Sloot, Eds., vol. 14074.\hskip 1em plus 0.5em minus 0.4em\relax Springer, 2023, pp. 93--107. [Online]. Available: \url{https://doi.org/10.1007/978-3-031-36021-3\_7}
\BIBentrySTDinterwordspacing

\bibitem{eskandari2019}
\BIBentryALTinterwordspacing
L.~Eskandari, J.~Mair, Z.~Huang, and D.~M. Eyers, ``I-scheduler: Iterative scheduling for distributed stream processing systems,'' \emph{Future Gener. Comput. Syst.}, vol. 117, pp. 219--233, 2021. [Online]. Available: \url{https://doi.org/10.1016/j.future.2020.11.011}
\BIBentrySTDinterwordspacing

\bibitem{ni2020}
\BIBentryALTinterwordspacing
X.~Ni, J.~Li, M.~Yu, W.~Zhou, and K.~Wu, ``Generalizable resource allocation in stream processing via deep reinforcement learning,'' in \emph{The Thirty-Fourth {AAAI} Conference on Artificial Intelligence, {AAAI} 2020, The Thirty-Second Innovative Applications of Artificial Intelligence Conference, {IAAI} 2020, The Tenth {AAAI} Symposium on Educational Advances in Artificial Intelligence, {EAAI} 2020, New York, NY, USA, February 7-12, 2020}.\hskip 1em plus 0.5em minus 0.4em\relax {AAAI} Press, 2020, pp. 857--864. [Online]. Available: \url{https://doi.org/10.1609/aaai.v34i01.5431}
\BIBentrySTDinterwordspacing

\bibitem{xu2014}
\BIBentryALTinterwordspacing
J.~Xu, Z.~Chen, J.~Tang, and S.~Su, ``T-storm: Traffic-aware online scheduling in storm,'' in \emph{{IEEE} 34th International Conference on Distributed Computing Systems, {ICDCS} 2014, Madrid, Spain, June 30 - July 3, 2014}.\hskip 1em plus 0.5em minus 0.4em\relax {IEEE} Computer Society, 2014, pp. 535--544. [Online]. Available: \url{https://doi.org/10.1109/ICDCS.2014.61}
\BIBentrySTDinterwordspacing

\bibitem{alnafessah2021}
\BIBentryALTinterwordspacing
A.~Alnafessah, G.~Russo~Russo, V.~Cardellini, G.~Casale, and F.~Lo~Presti, \emph{AI-Driven Performance Management in Data-Intensive Applications}.\hskip 1em plus 0.5em minus 0.4em\relax John Wiley \& Sons, Ltd, 2021, ch.~9, pp. 199--222. [Online]. Available: \url{https://onlinelibrary.wiley.com/doi/abs/10.1002/9781119675525.ch9}
\BIBentrySTDinterwordspacing

\bibitem{hilprecht2022}
\BIBentryALTinterwordspacing
B.~Hilprecht and C.~Binnig, ``Zero-shot cost models for out-of-the-box learned cost prediction,'' \emph{Proc. {VLDB} Endow.}, vol.~15, no.~11, pp. 2361--2374, 2022. [Online]. Available: \url{https://www.vldb.org/pvldb/vol15/p2361-hilprecht.pdf}
\BIBentrySTDinterwordspacing

\bibitem{ganapathi2009}
\BIBentryALTinterwordspacing
A.~Ganapathi, H.~A. Kuno, U.~Dayal, J.~L. Wiener, A.~Fox, M.~I. Jordan, and D.~A. Patterson, ``Predicting multiple metrics for queries: Better decisions enabled by machine learning,'' in \emph{Proceedings of the 25th International Conference on Data Engineering, {ICDE} 2009, March 29 2009 - April 2 2009, Shanghai, China}, Y.~E. Ioannidis, D.~L. Lee, and R.~T. Ng, Eds.\hskip 1em plus 0.5em minus 0.4em\relax {IEEE} Computer Society, 2009, pp. 592--603. [Online]. Available: \url{https://doi.org/10.1109/ICDE.2009.130}
\BIBentrySTDinterwordspacing

\bibitem{li2018}
\BIBentryALTinterwordspacing
T.~Li, Z.~Xu, J.~Tang, and Y.~Wang, ``Model-free control for distributed stream data processing using deep reinforcement learning,'' \emph{Proc. {VLDB} Endow.}, vol.~11, no.~6, pp. 705--718, 2018. [Online]. Available: \url{http://www.vldb.org/pvldb/vol11/p705-li.pdf}
\BIBentrySTDinterwordspacing

\bibitem{mao2019}
\BIBentryALTinterwordspacing
H.~Mao, M.~Schwarzkopf, S.~B. Venkatakrishnan, Z.~Meng, and M.~Alizadeh, ``Learning scheduling algorithms for data processing clusters,'' in \emph{Proceedings of the {ACM} Special Interest Group on Data Communication, {SIGCOMM} 2019, Beijing, China, August 19-23, 2019}, J.~Wu and W.~Hall, Eds.\hskip 1em plus 0.5em minus 0.4em\relax ACM, 2019, pp. 270--288. [Online]. Available: \url{https://doi.org/10.1145/3341302.3342080}
\BIBentrySTDinterwordspacing

\bibitem{hirzel2014}
\BIBentryALTinterwordspacing
M.~Hirzel, R.~Soul{\'{e}}, S.~Schneider, B.~Gedik, and R.~Grimm, ``A catalog of stream processing optimizations,'' \emph{{ACM} Comput. Surv.}, vol.~46, no.~4, pp. 46:1--46:34, 2013. [Online]. Available: \url{https://doi.org/10.1145/2528412}
\BIBentrySTDinterwordspacing

\bibitem{agnihotri2023}
\BIBentryALTinterwordspacing
P.~Agnihotri, B.~Koldehofe, C.~Binnig, and M.~Luthra, ``Zero-shot cost models for parallel stream processing,'' in \emph{Proceedings of the Sixth International Workshop on Exploiting Artificial Intelligence Techniques for Data Management, aiDM@SIGMOD 2023, Seattle, WA, USA, 18 June 2023}, R.~Bordawekar, O.~Shmueli, Y.~Amsterdamer, D.~Firmani, and A.~Kipf, Eds.\hskip 1em plus 0.5em minus 0.4em\relax ACM, 2023, pp. 5:1--5:5. [Online]. Available: \url{https://doi.org/10.1145/3593078.3593934}
\BIBentrySTDinterwordspacing

\bibitem{leis_2015}
\BIBentryALTinterwordspacing
V.~Leis, A.~Gubichev, A.~Mirchev, P.~A. Boncz, A.~Kemper, and T.~Neumann, ``How good are query optimizers, really?'' \emph{Proc. {VLDB} Endow.}, vol.~9, no.~3, pp. 204--215, 2015. [Online]. Available: \url{http://www.vldb.org/pvldb/vol9/p204-leis.pdf}
\BIBentrySTDinterwordspacing

\bibitem{siddiqui_2020}
\BIBentryALTinterwordspacing
T.~Siddiqui, A.~Jindal, S.~Qiao, H.~Patel, and W.~Le, ``Cost models for big data query processing: Learning, retrofitting, and our findings,'' in \emph{Proceedings of the 2020 International Conference on Management of Data, {SIGMOD} Conference 2020, online conference [Portland, OR, USA], June 14-19, 2020}, D.~Maier, R.~Pottinger, A.~Doan, W.~Tan, A.~Alawini, and H.~Q. Ngo, Eds.\hskip 1em plus 0.5em minus 0.4em\relax {ACM}, 2020, pp. 99--113. [Online]. Available: \url{https://doi.org/10.1145/3318464.3380584}
\BIBentrySTDinterwordspacing

\bibitem{gilmer2017}
\BIBentryALTinterwordspacing
J.~Gilmer, S.~S. Schoenholz, P.~F. Riley, O.~Vinyals, and G.~E. Dahl, ``Neural message passing for quantum chemistry,'' in \emph{Proceedings of the 34th International Conference on Machine Learning, {ICML} 2017, Sydney, NSW, Australia, 6-11 August 2017}, ser. Proceedings of Machine Learning Research, D.~Precup and Y.~W. Teh, Eds., vol.~70.\hskip 1em plus 0.5em minus 0.4em\relax PMLR, 2017, pp. 1263--1272. [Online]. Available: \url{http://proceedings.mlr.press/v70/gilmer17a.html}
\BIBentrySTDinterwordspacing

\bibitem{zaheer2017}
\BIBentryALTinterwordspacing
M.~Zaheer, S.~Kottur, S.~Ravanbakhsh, B.~P{\'{o}}czos, R.~Salakhutdinov, and A.~J. Smola, ``Deep sets,'' in \emph{Advances in Neural Information Processing Systems 30: Annual Conference on Neural Information Processing Systems 2017, December 4-9, 2017, Long Beach, CA, {USA}}, I.~Guyon, U.~von Luxburg, S.~Bengio, H.~M. Wallach, R.~Fergus, S.~V.~N. Vishwanathan, and R.~Garnett, Eds., 2017, pp. 3391--3401. [Online]. Available: \url{https://proceedings.neurips.cc/paper/2017/hash/f22e4747da1aa27e363d86d40ff442fe-Abstract.html}
\BIBentrySTDinterwordspacing

\bibitem{karimov2018}
\BIBentryALTinterwordspacing
J.~Karimov, T.~Rabl, A.~Katsifodimos, R.~Samarev, H.~Heiskanen, and V.~Markl, ``Benchmarking distributed stream data processing systems,'' in \emph{34th {IEEE} International Conference on Data Engineering, {ICDE} 2018, Paris, France, April 16-19, 2018}.\hskip 1em plus 0.5em minus 0.4em\relax {IEEE} Computer Society, 2018, pp. 1507--1518. [Online]. Available: \url{https://doi.org/10.1109/ICDE.2018.00169}
\BIBentrySTDinterwordspacing

\bibitem{akidau2015}
\BIBentryALTinterwordspacing
T.~Akidau, R.~Bradshaw, C.~Chambers, S.~Chernyak, R.~Fern{\'{a}}ndez{-}Moctezuma, R.~Lax, S.~McVeety, D.~Mills, F.~Perry, E.~Schmidt, and S.~Whittle, ``The dataflow model: {A} practical approach to balancing correctness, latency, and cost in massive-scale, unbounded, out-of-order data processing,'' \emph{Proc. {VLDB} Endow.}, vol.~8, no.~12, pp. 1792--1803, 2015. [Online]. Available: \url{http://www.vldb.org/pvldb/vol8/p1792-Akidau.pdf}
\BIBentrySTDinterwordspacing

\bibitem{chen2017}
\BIBentryALTinterwordspacing
X.~Chen, Y.~Vigfusson, D.~M. Blough, F.~Zheng, K.~Wu, and L.~Hu, ``{GOVERNOR:} smoother stream processing through smarter backpressure,'' in \emph{2017 {IEEE} International Conference on Autonomic Computing, {ICAC} 2017, Columbus, OH, USA, July 17-21, 2017}, X.~Wang, C.~Stewart, and H.~Lei, Eds.\hskip 1em plus 0.5em minus 0.4em\relax {IEEE} Computer Society, 2017, pp. 145--154. [Online]. Available: \url{https://doi.org/10.1109/ICAC.2017.31}
\BIBentrySTDinterwordspacing

\bibitem{carbone2015}
\BIBentryALTinterwordspacing
P.~Carbone, A.~Katsifodimos, S.~Ewen, V.~Markl, S.~Haridi, and K.~Tzoumas, ``Apache flink{\texttrademark}: Stream and batch processing in a single engine,'' \emph{{IEEE} Data Eng. Bull.}, vol.~38, no.~4, pp. 28--38, 2015. [Online]. Available: \url{http://sites.computer.org/debull/A15dec/p28.pdf}
\BIBentrySTDinterwordspacing

\bibitem{toshnival2014}
\BIBentryALTinterwordspacing
A.~Toshniwal, S.~Taneja, A.~Shukla, K.~Ramasamy, J.~M. Patel, S.~Kulkarni, J.~Jackson, K.~Gade, M.~Fu, J.~Donham, N.~Bhagat, S.~Mittal, and D.~V. Ryaboy, ``Storm@twitter,'' in \emph{International Conference on Management of Data, {SIGMOD} 2014, Snowbird, UT, USA, June 22-27, 2014}, C.~E. Dyreson, F.~Li, and M.~T. {\"{O}}zsu, Eds.\hskip 1em plus 0.5em minus 0.4em\relax ACM, 2014, pp. 147--156. [Online]. Available: \url{https://doi.org/10.1145/2588555.2595641}
\BIBentrySTDinterwordspacing

\bibitem{kulkarni2015}
\BIBentryALTinterwordspacing
S.~Kulkarni, N.~Bhagat, M.~Fu, V.~Kedigehalli, C.~Kellogg, S.~Mittal, J.~M. Patel, K.~Ramasamy, and S.~Taneja, ``Twitter heron: Stream processing at scale,'' in \emph{Proceedings of the 2015 {ACM} {SIGMOD} International Conference on Management of Data, Melbourne, Victoria, Australia, May 31 - June 4, 2015}, T.~K. Sellis, S.~B. Davidson, and Z.~G. Ives, Eds.\hskip 1em plus 0.5em minus 0.4em\relax ACM, 2015, pp. 239--250. [Online]. Available: \url{https://doi.org/10.1145/2723372.2742788}
\BIBentrySTDinterwordspacing

\bibitem{dutt2019}
\BIBentryALTinterwordspacing
A.~Dutt, C.~Wang, A.~Nazi, S.~Kandula, V.~Narasayya, and S.~Chaudhuri, ``Selectivity estimation for range predicates using lightweight models,'' \emph{Proc. VLDB Endow.}, vol.~12, no.~9, p. 1044–1057, may 2019. [Online]. Available: \url{https://doi.org/10.14778/3329772.3329780}
\BIBentrySTDinterwordspacing

\bibitem{chaudhary2020}
\BIBentryALTinterwordspacing
A.~Chaudhary, S.~Zeuch, and V.~Markl, ``Governor: Operator placement for a unified fog-cloud environment,'' in \emph{Proceedings of the 23rd International Conference on Extending Database Technology, {EDBT} 2020, Copenhagen, Denmark, March 30 - April 02, 2020}, A.~Bonifati, Y.~Zhou, M.~A.~V. Salles, A.~B{\"{o}}hm, D.~Olteanu, G.~H.~L. Fletcher, A.~Khan, and B.~Yang, Eds.\hskip 1em plus 0.5em minus 0.4em\relax OpenProceedings.org, 2020, pp. 631--634. [Online]. Available: \url{https://doi.org/10.5441/002/edbt.2020.81}
\BIBentrySTDinterwordspacing

\bibitem{kreps2011}
J.~Kreps, N.~Narkhede, J.~Rao \emph{et~al.}, ``Kafka: A distributed messaging system for log processing,'' in \emph{Proceedings of the NetDB}, vol.~11, no. 2011.\hskip 1em plus 0.5em minus 0.4em\relax Athens, Greece, 2011, pp. 1--7.

\bibitem{lightgbm}
\BIBentryALTinterwordspacing
G.~Ke, Q.~Meng, T.~Finley, T.~Wang, W.~Chen, W.~Ma, Q.~Ye, and T.~Liu, ``Lightgbm: {A} highly efficient gradient boosting decision tree,'' in \emph{Advances in Neural Information Processing Systems 30: Annual Conference on Neural Information Processing Systems 2017, December 4-9, 2017, Long Beach, CA, {USA}}, I.~Guyon, U.~von Luxburg, S.~Bengio, H.~M. Wallach, R.~Fergus, S.~V.~N. Vishwanathan, and R.~Garnett, Eds., 2017, pp. 3146--3154. [Online]. Available: \url{https://proceedings.neurips.cc/paper/2017/hash/6449f44a102fde848669bdd9eb6b76fa-Abstract.html}
\BIBentrySTDinterwordspacing

\bibitem{duplyakin2019}
\BIBentryALTinterwordspacing
D.~Duplyakin, R.~Ricci, A.~Maricq, G.~Wong, J.~Duerig, E.~Eide, L.~Stoller, M.~Hibler, D.~Johnson, K.~Webb, A.~Akella, K.~Wang, G.~Ricart, L.~Landweber, C.~Elliott, M.~Zink, E.~Cecchet, S.~Kar, and P.~Mishra, ``The design and operation of cloudlab,'' in \emph{2019 {USENIX} Annual Technical Conference, {USENIX} {ATC} 2019, Renton, WA, USA, July 10-12, 2019}, D.~Malkhi and D.~Tsafrir, Eds.\hskip 1em plus 0.5em minus 0.4em\relax {USENIX} Association, 2019, pp. 1--14. [Online]. Available: \url{https://www.usenix.org/conference/atc19/presentation/duplyakin}
\BIBentrySTDinterwordspacing

\bibitem{bordin_dspbench_2020}
\BIBentryALTinterwordspacing
M.~V. Bordin, D.~Griebler, G.~Mencagli, C.~F.~R. Geyer, and L.~G.~L. Fernandes, ``Dspbench: {A} suite of benchmark applications for distributed data stream processing systems,'' \emph{{IEEE} Access}, vol.~8, pp. 222\,900--222\,917, 2020. [Online]. Available: \url{https://doi.org/10.1109/ACCESS.2020.3043948}
\BIBentrySTDinterwordspacing

\bibitem{hesse_espbench_2021}
\BIBentryALTinterwordspacing
G.~Hesse, C.~Matthies, M.~Perscheid, M.~Uflacker, and H.~Plattner, ``Espbench: The enterprise stream processing benchmark,'' in \emph{{ICPE} '21: {ACM/SPEC} International Conference on Performance Engineering, Virtual Event, France, April 19-21, 2021}, J.~Bourcier, Z.~M.~J. Jiang, C.~Bezemer, V.~Cortellessa, D.~D. Pompeo, and A.~L. Varbanescu, Eds.\hskip 1em plus 0.5em minus 0.4em\relax ACM, 2021, pp. 201--212. [Online]. Available: \url{https://doi.org/10.1145/3427921.3450242}
\BIBentrySTDinterwordspacing

\bibitem{shukla2017}
\BIBentryALTinterwordspacing
A.~Shukla, S.~Chaturvedi, and Y.~Simmhan, ``Riotbench: An iot benchmark for distributed stream processing systems,'' \emph{Concurr. Comput. Pract. Exp.}, vol.~29, no.~21, 2017. [Online]. Available: \url{https://doi.org/10.1002/cpe.4257}
\BIBentrySTDinterwordspacing

\bibitem{lu2014}
\BIBentryALTinterwordspacing
R.~Lu, G.~Wu, B.~Xie, and J.~Hu, ``Stream bench: Towards benchmarking modern distributed stream computing frameworks,'' in \emph{Proceedings of the 7th {IEEE/ACM} International Conference on Utility and Cloud Computing, {UCC} 2014, London, United Kingdom, December 8-11, 2014}.\hskip 1em plus 0.5em minus 0.4em\relax {IEEE} Computer Society, 2014, pp. 69--78. [Online]. Available: \url{https://doi.org/10.1109/UCC.2014.15}
\BIBentrySTDinterwordspacing

\bibitem{jerzak2014}
\BIBentryALTinterwordspacing
Z.~Jerzak and H.~Ziekow, ``The {DEBS} 2014 grand challenge,'' in \emph{The 8th {ACM} International Conference on Distributed Event-Based Systems, {DEBS} '14, Mumbai, India, May 26-29, 2014}, U.~Bellur and R.~Kothari, Eds.\hskip 1em plus 0.5em minus 0.4em\relax ACM, 2014, pp. 266--269. [Online]. Available: \url{https://doi.org/10.1145/2611286.2611333}
\BIBentrySTDinterwordspacing

\bibitem{koliousis2016}
\BIBentryALTinterwordspacing
A.~Koliousis, M.~Weidlich, R.~C. Fernandez, A.~L. Wolf, P.~Costa, and P.~R. Pietzuch, ``{SABER:} window-based hybrid stream processing for heterogeneous architectures,'' in \emph{Proceedings of the 2016 International Conference on Management of Data, {SIGMOD} Conference 2016, San Francisco, CA, USA, June 26 - July 01, 2016}, F.~{\"{O}}zcan, G.~Koutrika, and S.~Madden, Eds.\hskip 1em plus 0.5em minus 0.4em\relax ACM, 2016, pp. 555--569. [Online]. Available: \url{https://doi.org/10.1145/2882903.2882906}
\BIBentrySTDinterwordspacing

\bibitem{peng2015}
\BIBentryALTinterwordspacing
B.~Peng, M.~Hosseini, Z.~Hong, R.~Farivar, and R.~H. Campbell, ``R-storm: Resource-aware scheduling in storm,'' in \emph{Proceedings of the 16th Annual Middleware Conference, Vancouver, BC, Canada, December 07 - 11, 2015}, R.~Lea, S.~Gopalakrishnan, E.~Tilevich, A.~L. Murphy, and M.~Blackstock, Eds.\hskip 1em plus 0.5em minus 0.4em\relax ACM, 2015, pp. 149--161. [Online]. Available: \url{https://doi.org/10.1145/2814576.2814808}
\BIBentrySTDinterwordspacing

\bibitem{alexandrov2014}
\BIBentryALTinterwordspacing
A.~Alexandrov, R.~Bergmann, S.~Ewen, J.~Freytag, F.~Hueske, A.~Heise, O.~Kao, M.~Leich, U.~Leser, V.~Markl, F.~Naumann, M.~Peters, A.~Rheinl{\"{a}}nder, M.~J. Sax, S.~Schelter, M.~H{\"{o}}ger, K.~Tzoumas, and D.~Warneke, ``The stratosphere platform for big data analytics,'' \emph{{VLDB} J.}, vol.~23, no.~6, pp. 939--964, 2014. [Online]. Available: \url{https://doi.org/10.1007/s00778-014-0357-y}
\BIBentrySTDinterwordspacing

\bibitem{foroni2018}
\BIBentryALTinterwordspacing
D.~Foroni, C.~Axenie, S.~Bortoli, M.~A.~H. Hassan, R.~Acker, R.~Tudoran, G.~Brasche, and Y.~Velegrakis, ``Moira: {A} goal-oriented incremental machine learning approach to dynamic resource cost estimation in distributed stream processing systems,'' in \emph{Proceedings of the International Workshop on Real-Time Business Intelligence and Analytics, {BIRTE} 2018, Rio de Janeiro, Brazil, August 27, 2018}, M.~Castellanos, P.~K. Chrysanthis, B.~Chandramouli, and S.~Chen, Eds.\hskip 1em plus 0.5em minus 0.4em\relax ACM, 2018, pp. 2:1--2:10. [Online]. Available: \url{https://doi.org/10.1145/3242153.3242160}
\BIBentrySTDinterwordspacing

\bibitem{li2016}
\BIBentryALTinterwordspacing
T.~Li, J.~Tang, and J.~Xu, ``Performance modeling and predictive scheduling for distributed stream data processing,'' \emph{{IEEE} Trans. Big Data}, vol.~2, no.~4, pp. 353--364, 2016. [Online]. Available: \url{https://doi.org/10.1109/TBDATA.2016.2616148}
\BIBentrySTDinterwordspacing

\bibitem{luthra2018}
\BIBentryALTinterwordspacing
M.~Luthra, B.~Koldehofe, P.~Weisenburger, G.~Salvaneschi, and R.~Arif, ``{TCEP:} adapting to dynamic user environments by enabling transitions between operator placement mechanisms,'' in \emph{Proceedings of the 12th {ACM} International Conference on Distributed and Event-based Systems, {DEBS} 2018, Hamilton, New Zealand, June 25-29, 2018}, A.~Hinze, D.~M. Eyers, M.~Hirzel, M.~Weidlich, and S.~Bhowmik, Eds.\hskip 1em plus 0.5em minus 0.4em\relax ACM, 2018, pp. 136--147. [Online]. Available: \url{https://doi.org/10.1145/3210284.3210292}
\BIBentrySTDinterwordspacing

\end{thebibliography}
\end{document}